\documentclass[12pt,a4paper]{article}
\pdfoutput=1

\usepackage[english]{babel} 
\usepackage[utf8]{inputenc}
\usepackage[left=01in,right=1in,top=1.0in,bottom=1.1in]{geometry}
\usepackage[labelfont=bf]{caption}
\usepackage[hyperfootnotes=false]{hyperref}
\hypersetup{
 pdfauthor={Rafael Aoude, Gilly Elor, Grant N. Remmen, Olcyr Sumensari}}
\usepackage[normalem]{ulem}

\usepackage{bbm}
\usepackage{bm}

\usepackage{dsfont}
\usepackage{fontawesome5}
\usepackage{amsfonts}
\usepackage{amsmath}
\usepackage{amssymb}
\usepackage{amsthm}
\usepackage{bbold}
\usepackage{calc}
\usepackage{cancel}
\usepackage{float}
\usepackage{slashed}
\usepackage{mathrsfs} 
\usepackage{pdfpages}
\usepackage{multirow}
\usepackage{mathtools}
\usepackage{braket}
\usepackage{cite}
\usepackage{tikz}
\usepackage{tikzsymbols}
\usepackage{caption}
\usepackage{subcaption}
\usepackage{array}
\usepackage{colortbl} 	
\definecolor{Gray}{gray}{0.95}
\definecolor{RGray}{gray}{0.90}
\definecolor{CGray}{gray}{0.92}
\usepackage{arydshln}
\usepackage{soul}
\usepackage{booktabs}

\usepackage[hang,flushmargin]{footmisc}
\usepackage{scalerel}
\usepackage{xcolor}

\definecolor{codegreen}{rgb}{0,0.6,0}
\definecolor{codegray}{rgb}{0.5,0.5,0.5}
\definecolor{codepurple}{rgb}{0.58,0,0.82}
\definecolor{backcolour}{rgb}{0.95,0.95,0.92}
\definecolor{lightgray}{rgb}{0.9,0.9,0.9}
\definecolor{niceblue}{rgb}{0.15,0.15,0.6}
\definecolor{nicegreen}{rgb}{0.1,0.5,0.1}
\definecolor{Red}{rgb}{1.,0.,0.}

\definecolor{Green}{rgb}{0.2,.7,0.2}

\usepackage{listings}
\lstdefinestyle{mystyle}{
    backgroundcolor=\color{backcolour},   
    commentstyle=\color{codegreen},  
    keywordstyle=\color{magenta},
    numberstyle=\tiny\color{codegray},
    stringstyle=\color{codepurple},
    basicstyle=\ttfamily\footnotesize,
    breakatwhitespace=false,         
    breaklines=true,                 
    captionpos=b,                    
    keepspaces=true,                 
    numbers=left,                    
    numbersep=5pt,                  
    showspaces=false,                
    showstringspaces=false,
    showtabs=false,                  
    tabsize=2
}
\hypersetup{
    colorlinks=true,
    linkcolor=niceblue,
    citecolor=cadmiumgreen,
    filecolor=cadmiumgreen,      
    urlcolor=cadmiumgreen,
    linktoc=all
}
\usepackage{lipsum}
\numberwithin{equation}{section}
\numberwithin{figure}{section}
\numberwithin{table}{section}

\renewcommand\thefootnote{\textcolor{cadmiumgreen}{\arabic{footnote}}}

\allowdisplaybreaks

\usetikzlibrary{decorations}

\pgfdeclaredecoration{complete sines}{initial}
{
    \state{initial}[
        width=+0pt,
        next state=sine,
        persistent precomputation={\pgfmathsetmacro\matchinglength{
            \pgfdecoratedinputsegmentlength / int(\pgfdecoratedinputsegmentlength/\pgfdecorationsegmentlength)}
            \setlength{\pgfdecorationsegmentlength}{\matchinglength pt}
        }] {}
    \state{sine}[width=\pgfdecorationsegmentlength]{
        \pgfpathsine{\pgfpoint{0.25\pgfdecorationsegmentlength}{0.5\pgfdecorationsegmentamplitude}}
        \pgfpathcosine{\pgfpoint{0.25\pgfdecorationsegmentlength}{-0.5\pgfdecorationsegmentamplitude}}
        \pgfpathsine{\pgfpoint{0.25\pgfdecorationsegmentlength}{-0.5\pgfdecorationsegmentamplitude}}
        \pgfpathcosine{\pgfpoint{0.25\pgfdecorationsegmentlength}{0.5\pgfdecorationsegmentamplitude}}
}
    \state{final}{}
}

\tikzset{vector/.style={decorate, decoration={complete sines, amplitude=8pt, segment length=10pt}}}

\tikzset{
wc/.style = {circle, fill, minimum size=#1,
              inner sep=0pt, outer sep=0pt},
wc/.default = 6pt 
}

\graphicspath{{./Figures/}{./figures/}}

\newcommand{\be}{\begin{equation}}
\newcommand{\ee}{\end{equation}}
\newcommand{\bea}{\begin{eqnarray}}
\newcommand{\eea}{\end{eqnarray}}

\newcommand{\rma}{{\rm a}}

\newcommand{\rin}{{\rm in}}

\newcommand{\Fn}{{N}}

\definecolor{cadmiumgreen}{rgb}{0.0, 0.42, 0.24}

\definecolor{applegreen}{rgb}{0.55, 0.71, 0.0}

\newcommand\blfootnote[1]{%
  \begingroup
  \renewcommand\thefootnote{}\footnote{#1}%
  \addtocounter{footnote}{-1}%
  \endgroup
}
\makeatletter
\g@addto@macro\bfseries{\boldmath}
\makeatother

\newcommand{\cA}{{\mathcal A}}

\begin{document}
 \begin{flushright}
\,\\
 \end{flushright}

\begin{center}
{\Large\bf Positivity in Amplitudes and Quantum
Entanglement}
\\[1.0cm]
{\sc Rafael Aoude$^{\,a}$\blfootnote{\href{mailto:eboli@if.usp.br}{rafael.aoude@ed.ac.uk}}, 
Gilly Elor$^{\,b}$\blfootnote{\href{mailto:luighi.leal@usp.br}{gilly.elor@austin.utexas.edu}},
Grant N. Remmen$^{\,c}$\blfootnote{\href{mailto:matheus.martines.silva@usp.br}{grant.remmen@nyu.edu}},
Olcyr~Sumensari$^{\,d}$\blfootnote{\href{mailto:olcyr.sumensari@ijclab.in2p3.fr}{olcyr.sumensari@ijclab.in2p3.fr}}}
\vspace{0.7cm}

{\em\small ${}^{\ a}$Higgs Centre for Theoretical Physics, School of Physics and Astronomy \\ The University of Edinburgh, Edinburgh EH9 3JZ, Scotland, United Kingdom}\\[0.2em]
{\em \small${}^{\ b}$Weinberg Institute, Department of Physics, University of Texas at Austin, Austin, TX 78712,
United States}\\[0.2em]
{\em\small ${}^{\ c}$Center for Cosmology and Particle Physics, Department of Physics \\
New York University, New York, NY 10003, United States}\\[0.2em]
{\em \small${}^{\ d}$IJCLab, P\^ole Th\'eorie (Bât.~210), CNRS/IN2P3 et Univ.~Paris-Saclay, 91405 Orsay, France}\\[0.2em]
\end{center}
\vspace{0.5 cm}

\centerline{\large\bf Abstract}
\begin{quote}
We explore the connection of positivity of the imaginary part of forward elastic amplitudes for perturbative scattering with consistency of the entanglement generated by the S-matrix, for states with arbitrary internal quantum numbers such as flavor. We also analyze ``disentanglers,'' certain highly entangled initial states for which the action of the S-matrix is to decrease subsystem entanglement. As a by-product, we develop a framework based on wave packets to regularize the spacetime divergences that appear in the plane-wave derivation of the $2\to2$ entanglement expression.
\end{quote}
\thispagestyle{empty}
\clearpage
\setcounter{page}{1}

 \clearpage



\section{Introduction}Scattering amplitudes are the fundamental objects of study in quantum field theory, underpinning much of modern high-energy physics, from phenomenology to string theory.
The mathematical structure of amplitudes has been studied for decades, and important theorems arising from unitarity and locality are well known, constraining their asymptotic growth and analytic structure~\cite{froissart1961asymptotic,PhysRev.129.1432,Lukaszuk:1967zz}.
Moreover, in the modern amplitudes program, it has been shown that unitarity and locality themselves can emerge from more primordial mathematical structures~\cite{Arkani-Hamed:2016rak,Arkani-Hamed:2023lbd}.

Chief among the many famous facts about scattering amplitudes is the optical theorem, which relates the imaginary part of an amplitude---at forward kinematics, where the outgoing and incoming states are identical and particles pass straight through each other---to its cross section.
Systematic use of this ``positivity'' of amplitudes has especially in recent years allowed the laws of physics themselves to be constrained using  dispersion relations, implying numerous different bounds on the space of coefficients of higher-dimension operators in effective field theories (EFTs)~\cite{Adams:2006sv,Jenkins:2006ia,Nicolis:2009qm,Dvali:2012zc,Arkani-Hamed:2020blm,Sinha:2020win,Cheung:2025nhw}, ranging from the standard model EFT~\cite{Remmen:2019cyz,Remmen:2020vts,Remmen:2020uze,Remmen:2022orj,Zhang:2020jyn,Trott:2020ebl,Remmen:2024hry} to quantum corrections to gravity~\cite{Bellazzini:2015cra,Cheung:2016wjt,Cheung:2014ega,Hamada:2018dde,Arkani-Hamed:2021ajd}. 

Meanwhile, entanglement is the quintessential  feature of quantum mechanics~\cite{Einstein:1935rr,Bell:1964kc}. Experimentally verified in classic demonstrations of Bell's inequality violations~\cite{Bell:1964kc,CHSH_1969}, the highest-energy probes of entanglement can now be implemented in colliders, such as  measurements of spin entanglement in top quark systems recently observed by the ATLAS Collaboration at the LHC~\cite{ATLAS:2023fsd} as proposed in \cite{Afik:2020onf}; see also Refs.~\cite{Barr:2021zcp,Severi:2021cnj,Aguilar-Saavedra:2022uye,Ashby-Pickering:2022umy,Sinha:2022crx,Aoude:2022imd,Fabbrichesi:2022ovb,Severi:2022qjy}.

Quantum information theory also provides many positivity bounds on physics of an a priori different sort than dispersion relations, taking the form of inequalities on the von~Neumann entropy $S(\hat \rho) = -{\rm Tr}[\hat \rho\log \hat \rho]$ for a density matrix $\hat{\rho}$, including monogamy of entanglement, strong subadditivity,  etc.~\cite{nielsen00}.
By considering other quantum information theoretic measures, such as relative entropy, yet more relations can be found~\cite{Lindblad}; see also ~\cite{Hayden:2011ag,Dutta:2019gen,Bao:2021vyq} for further bounds derived in holographic contexts.
Underpinning all of these results is the basic requirement of positivity of entanglement entropy, $S(\hat \rho)\geq 0$.

An interesting question is whether these two notions of positivity---that is, the amplitudes' positivity and entanglement entropy---can be related.
Do quantum information bounds place additional constraints on EFTs? Or is the optical theorem itself a statement about entanglement in the S-matrix?
{ The intersection of quantum entanglement and scattering is a subject of increasing interest~\cite{Low:2024hvn,Kowalska:2024kbs,Kowalska:2025qmf,Peschanski:2026edo,Cheung:2023hkq}, including its relation to emergent symmetries~\cite{Beane:2018oxh,Low:2021ufv,Aoude:2020mlg,Chen:2021huj,Carena:2023vjc}, as well as the Yang Mills/gravity double copy~\cite{Cheung:2020uts}.}
Notably, Ref.~\cite{Cheung:2023hkq} considered the quantum Tsallis entropy and showed that, subject to certain conditions on an unentangled initial state, it grows under perturbative scattering.
The question of whether positivity of the quantum entanglement---computed formally as a mathematical operation on the object obtained by evolving the state with the S-matrix, irrespective of whether that S-matrix is unitarity---itself can {\it imply} amplitudes' positivity, in a broader context of states with general quantum numbers, especially relevant for both the contexts of quantum computing and particle phenomenology, has not been explored.

In this paper, we address this problem and show that amplitudes' positivity is related to positivity of another object, specifically the trace of the square of the state acted upon with the S-matrix, for a general perturbative quantum theory.
That is, we define the following quantity ${\cal E}$, computed from the state via
\begin{equation}
\begin{aligned}
{\cal E}[\Omega] &\equiv 1- \text{Tr}_A \left[ \hat{\rho}_A^2 \right] \,, \qquad \hat{\rho}_A \equiv \text{Tr}_B \hat{\rho}_{AB}\,, \qquad \hat{\rho}_{AB} \equiv \hat{S} |\Omega \rangle \langle \Omega | \hat{S}^\dagger\,,
\label{eq:linE}
\end{aligned}
\end{equation}
for a two-to-two scattering process through S-matrix $\hat{S}$. For a unitary S-matrix, ${\cal E}$ is the ``linearized entropy,'' and is by definition positive. However, we would like to investigate the converse, namely, without assuming that $\hat S$ is unitary, whether merely {\it imposing} positivity of ${\cal E}$ implies consequences for the dynamics. 
{While of course ${\cal E} \geq 0$ in a unitary theory, by virtue of $\hat\rho_A$
 being a well-defined density matrix, here we are instead interested in formally computing $\cal E$ via perturbative scattering and checking what the mechanical consequences of its positivity are at the level of the amplitude.} We will find that $\cal E$ is nonnegative {\it if and only if} the corresponding amplitude has positive imaginary part in the forward limit, i.e.,
\be
\mathcal{E}\left[|\Omega^{\rm prod} \rangle \right] \geq 0 \quad \Longleftrightarrow \quad  \text{Im} \,\mathcal{A} > 0 \,.
\label{eq:goal}
\ee
We reiterate that, the object ${\cal E}$ is only truly a well defined entropy in the usual sense if the S-matrix is unitary; however, for the purposes of this paper, we will take ${\cal E}$ to be the mathematical object defined in Eq.~\eqref{eq:linE}, irrespective of $\hat \rho_A$ and $\hat S$.
{Again, we emphasize that we are simply going to compute ${\cal E}$ as a perturbatively defined object at the level of the scattering amplitude, demand positivity, and check what this implies about the amplitude itself. While unitarity of $\hat S$ implies positivity of ${\cal E}$, so the implication $\impliedby$ in Eq.~\eqref{eq:goal} is immediate, the converse ($\implies$) will be interesting to show.}
That is, we will show that the elastic optical theorem implies $\mathcal{E}\left[|\Omega^{\rm prod} \rangle \right] \geq 0$.
We will write a general initial two-particle state in Hilbert space $\mathcal{H}_A \otimes \mathcal{H}_B$ 
as $ |\Omega \rangle = \sum_{ab} \Omega_{ab} |k_1,a\rangle_A \otimes |k_2,b \rangle_B$, 
with definite momenta $k_i$ and summed over internal quantum numbers, and $\Omega^{\rm prod}$ refers to a product (unentangled) initial state. For the basis vectors we use the notation $|k_1,a\rangle_A \otimes |k_2,b \rangle_B \equiv |k_1,a;k_2,b \rangle$, and write $\hat{\rho}_{A}$ for the reduced density matrix in $\mathcal{H}_A$ after tracing out degrees of freedom in $\mathcal{H}_B$.
The results of this work represent an exciting connection between entanglement and the S-matrix, opening the door to future insights that might be gained in the amplitudes program via the use of powerful results from quantum information theory. From a phenomenological perspective, the study of entanglement in EFTs could yield insights to model building principles in a similar spirit to  Refs.~\cite{Beane:2018oxh,Low:2021ufv,Aoude:2020mlg,Chen:2021huj,Carena:2023vjc}.
\bigskip

\section{Entanglement from Two-to-Two Scattering}
\label{sec:ent-2to2}

The entanglement entropy can be computed perturbatively, order by order, in a quantum field theory \cite{Rosenhaus_2014}. Furthermore, such calculations have been shown to be valid in a Wilsonian EFT~\cite{Balasubramanian:2011wt, Costa:2022bvs} where one deals with effective operators in the infrared. In practice, this amounts to tracing over continuous degrees of freedom. For example, Refs.~\cite{Seki:2014cgq,Peschanski,Peschanski:2019yah,Fan:2017hcd} built upon the results of Ref.~\cite{Balasubramanian:2011wt} to compute momentum-space entanglement generated in scattering (with and without spin entanglement). In this paper, we are interested in the quantity ${\cal E}$ defined in Eq.~\eqref{eq:linE} generated between scalar fields with an internal quantum number, e.g., flavor; we will address the generalization to fermions or vectors in future work. We thus compute the traces of operators $\hat{\mathcal{O}}$ as 
\bea
\text{Tr}_B [ \hat{\mathcal{O}} ] =  \sum_c \int_u  \langle u, c |_B \hat{\mathcal{O}} |u,c \rangle_B\,,
\eea 
We use the shorthand notation for the phase-space integral $\int_p  \equiv \int d^3 p / [(2\pi)^3 2 E_p]$ throughout the paper. The reduced density matrix in Eq.~\eqref{eq:linE} of a scalar particle in $\mathcal{H}_A$ is computed by tracing out degrees of freedom in $\mathcal{H}_B$,
\be
\hat{\rho}_A  =\frac{1}{\Fn}  \sum_c \int_u \, \langle u, c |_B   \hat{S} |\Omega \rangle \langle \Omega | \hat{S}^\dagger |  u,c \rangle_B \,.
\label{eq:rhoAintro}
\ee
Here $\Fn$ is a factor that ensures that the initial states (for which $\hat{S}$ is replaced by $\hat{\mathbbm{1}}$) are properly normalized~\footnote{For a general state the normalization $N$ is given by the integral of $\mathcal{F}(k,k)_{aa}$ over $k$, summed over $a$, where $\mathcal{F}$ is defined in Eq.~\eqref{eq:scalarfunc}.}. 
We now compute a quantity that, in a unitary theory, corresponds to the linearized entanglement generated by two-to-two scalar scattering $\phi_i(k_1)\phi_j (k_2) \rightarrow \phi_m(p_1) \phi_n (p_2)$, induced by a perturbative S-matrix $\hat{S} = \hat{\mathbbm{1}} + i \hat{T}$ but \emph{without assuming unitarity}. The indices $i,j,m,n$ denote internal quantum numbers and $p/k$ are outgoing/incoming momentum.  This computation holds both for the case where scattering proceeds through a four-scalar contact term or where the scattering in generated by a mediating particle.

As stated above we will \emph{not} assume unitarity. Our assumption is simply that there exists an S-matrix, which we will take to be a {\it state-independent general linear operator} acting on the Hilbert space.
We will {\it not} assume a priori that $\hat S^\dagger \hat S = \hat{\mathbbm{1}}$~\footnote{Here, $\hat S^\dagger$ is defined as the Hermitian conjugate of $\hat S$, i.e., where $\hat S$ acts with left multiplication on vectors $|\psi\rangle$ in Hilbert space, $\hat S^\dagger$ acts with right multiplication on covectors $\langle\psi|$.}.
Perturbative scattering amplitudes are defined by acting on state vectors with the transition matrix $\hat{T}$, i.e., $\langle p_1,m; p_2,n |  \hat{T}| k_1, a; k_2,b \rangle = (2\pi)^4 \delta^4 \left( k_{i} - p_{f} \right) \mathcal{A}_{ab}^{mn} (k_1 k_2 \rightarrow p_1 p_2 )$. 
While momentum conservation is enforced through the delta functions in the definitions above, \emph{we do not demand conservation of the internal symmetry}.
Taking the Hermitian conjugate of the S-matrix defines $\mathcal{A}^\dagger$, without assuming that it is the CPT conjugate of $\mathcal{A}$. Since we have not assumed unitarity, we also do not a priori have the optical theorem, and so the sign of ${\rm Im}\,{\cal A}$ is as yet unfixed. 
The two-particle incoming pure state $|\Omega \rangle$ is defined below Eq.~\eqref{eq:linE}. The internal quantum numbers of the incoming states live in $\mathcal{H}_A \otimes \mathcal{H}_B$, which we take as discrete, e.g., $\mathds{C}^n \otimes \mathds{C}^n$, so that we can use the above expressions to formally compute the entanglement generated in scattering. 
Notwithstanding the optical theorem, we will consider the quantity ${\cal E}$ in Eq.~\eqref{eq:linE}, corresponding to the trace of the square of the reduced state of one of the outgoing particles, irrespective of whether the S-matrix is unitary.
Of course, in a unitary theory, ${\cal E}$ corresponds to the linearized entropy and is by definition positive.
Positivity of entropy is itself obtained in quantum mechanics using properly normalized density matrices, with the preservation of unit normalization of the density matrix a consequence of unitarity; in this paper, we are interested in the logical converse, where we elevate positivity of {$\cal E$} to an axiom, and obtain the flagship consequence of unitarity, namely, positivity in the optical theorem sense, as an output rather than an input.
We wish to find the positivity implied by the optical theorem as a {\it consequence} of positivity of ${\cal E}$.

In the following, we provide a detailed derivation of the linearized quantity $\Delta \mathcal{E} [\Omega]$ defined in Eq.~\eqref{eq:linE}, which can be written as
\be
\mathcal{E}[\Omega] =  1 - \frac{1}{\Fn^2} \sum_{ab} \int_{q_1} \int_{q_2}  \mathcal{F}(q_1,q_2)_{ab} \mathcal{F}(q_2,q_1)_{ba}
\label{eq:entpower}
\ee
in terms of a scalar function,
\be
\hspace{-1mm} \mathcal{F}(q_1,q_2)_{ab}  \,{\equiv}\, \sum_m \! \int_l \,  \langle q_1,a;l,m | \hat{S} |\Omega \rangle \langle \Omega | \hat{S}^\dagger |q_2, b;l,m \rangle\,.
\label{eq:scalarfunc}
\ee
We emphasize that the $\hat S^\dagger$ appearing above does not arise from any assumption of unitarity, but instead simply from the definition of bras, kets, and the inner product. In Sec.~\ref{ssec:plane-wave} we show that the leading-order change in $\cal E$ of the two scalars generated by scattering is given by 
\be
\begin{aligned}
\label{eq:DeltaE}
&\Delta \mathcal{E}[\Omega ] =   \frac{4(2\pi)^4 \delta^{(4)}(0)}{N}  \,\text{Im} \Bigl[ \sum_{abij} \Omega_{ij}  \mathcal{A}^{ij}_{ab} (k_1 k_2 {\rightarrow} k_1 k_2 ) (\Omega^\dag \cdot \Omega \cdot \Omega^\dag)_{ab}  \Bigr]\,,
\end{aligned}
\ee
through $O(g^2)$, where $g$ is a small coupling in the perturbative expansion $\mathcal{A} \sim g^2$. Here $\Delta \mathcal{E} =\mathcal{E}_f - \mathcal{E}_i$ and $\mathcal{E}_i[\Omega ] = 1- \text{Tr} \left[ (\Omega^\dagger \Omega)^2 \right]$, which vanishes for a product initial state. 
Divergences in spacetime volume are defined as $V \equiv (2\pi)^3 \delta_V^3 (0)$  and  $T \equiv 2 \pi \delta_T (0)$. The seemingly divergent prefactor in Eq.~\eqref{eq:DeltaE}, which we henceforth denote as 
\begin{equation}
\label{eq:calN}
{\cal N} \equiv { \dfrac{(2\pi)^4 \delta^ {(4)}(0)}{N}=}\frac{1}{2 E_{k_1} 2 E_{k_2}} \frac{T}{V}  \,, 
\end{equation}
which is an artifact of using plane waves in our derivation. If we instead use wave packets, as in 
section~\ref{sec:wave-function-derivation}, we find that ${\cal N}$ is nondivergent.
For properly normalized initial states $\Omega$, with  $\text{Tr} \left[ \Omega^\dagger \Omega \right]  = 1 $, we have $\Fn = 2E_{k_1} 2 E_{k_2} V^2$.  Interestingly, the computation of ${\cal E}$ has automatically selected amplitudes with forward kinematics from the integral in Eq.~\eqref{eq:entpower}.

\subsection{Plane Wave Derivation}
\label{ssec:plane-wave}

We now present details of the computation of 
$\Delta \mathcal{E} [\Omega ]$ in Eq.~\eqref{eq:DeltaE} (which in a unitary theory is the change in the linearized entanglement) generated in 
$\phi_i (k_1) \phi_j (k_2) \rightarrow \phi_k (p_1) \phi_l (p_2)$ scattering. First, we will compute $\cal E$ for a general initial state $|\Omega \rangle$ using the formalism discussed in this paper. We then compute the leading-order $\cal E$ generated by the final state through the action of a perturbative S-matrix. We do this first in the case where the scattering proceeds through a four-point scalar contact term. We then repeat this computation in the case where the scattering arises through the exchange of a heavy particle. In both cases, we find that Eq.~\eqref{eq:DeltaE} describes a quantity that in a unitary theory is the leading-order linearized entanglement generated through the scattering, as must have been the case. We present our derivation in terms of a toy model, but the result holds for, e.g., photon scattering $\gamma \gamma \rightarrow \Phi \rightarrow \gamma \gamma$ mediated by a heavy scalar $\Phi$, where transverse photon helicities can be treated as an internal symmetry in the forward limit.

Throughout this derivation, our starting point will be a general two-particle incoming state,
 \be
 |\Omega \rangle = \sum_{ij} \Omega_{ij} |k_1,i\rangle_A  \otimes |k_2, j \rangle_B \equiv  \sum_{ij} \Omega_{ij} |k_1,i;k_2,j \rangle \,,
\label{eq:omegaapp}
 \ee
where $\phi_i \in \mathcal{H}_A$ and $\phi_j \in \mathcal{H}_B$. In this section, we are going to use plane-wave states $|k,a\rangle$, which have infinite normalization~\cite{Fan:2017hcd}
\be
\langle k,a | l,b \rangle = (2 \pi)^3 2 E_k \delta^3(k-l) \delta_{ab}
\ee
leading to spacelike and timelike divergences as $V \equiv (2\pi)^3 \delta_V^3 (0)$  and  $T \equiv (2 \pi) \delta_T (0)$. We will then compute $\cal E$ from Eq.~\eqref{eq:linE}, in the form $\hat{\mathbbm{1}} = \sum_n\hat{\mathbbm{1}}_n$, where the $n$-particle identity is given by $\hat{\mathbbm{1}}_n = (\prod_{i =1 }^{n} \int_{q_i} ) | \left\{ q_i \right\} \rangle \langle \left\{ q_i \right\} |$ 
 and we sum over $n$-particle intermediate states, $\left\{ q_i \right\} \equiv  \left\{ q_1, q_2 ,\ldots, q_n \right\}$\,.

\subsubsection{Entanglement of the initial state}

We begin with the straightforward derivation of the initial state entanglement. The full system density matrix of the two-particle initial state is constructed as follows:
\bea
\hat{\rho}^i_{AB} = |\Omega \rangle \langle \Omega | = \sum_{ijab} \Omega_{ij}\Omega_{ab}^\dag |k_1, i \rangle_A \otimes |k_2, j \rangle_B  \langle k_1,a |_A \otimes \langle k_2,b |_B\,.
\label{eq:Omegain}
\eea
The reduced density matrix is computed by tracing over particle $B$,
\be 
\begin{aligned}
\text{Tr}_B[\hat{\rho}_{AB}^i] 
&=2E_{k_2} V 
 \sum_{ija}\Omega_{ij}\Omega_{aj}^\dag | k_1,i \rangle_A \langle k_1,a |_A\,.
\end{aligned}
\ee
The initial density matrix must be properly normalized so that $\text{Tr}_A \hat{\rho}_{nA}^i = 1$, 
\begin{equation}
\begin{aligned}
\Fn = \text{Tr}_A [\text{Tr}_B \hat{\rho}^i_{AB}] &= 2E_{k_2} 2 E_{k_1}V^2 \text{Tr} [ \Omega^\dag \Omega] = 2E_{k_2} 2 E_{k_1}V^2\,.
\label{eq:Ein}
\end{aligned}
\end{equation}
The linear entanglement is then
\be 
\begin{aligned}
\mathcal{E}_i[\Omega] &= 1 - \frac{1}{\Fn^2} \text{Tr}_A [\hat{\rho}_A^2] 
    = 1 -  
\text{Tr}[(\Omega^\dag \Omega)^2] \,.
\label{eq:Elinscat}
\end{aligned}
\ee
Note that for a pure initial state $\text{Tr}[(\Omega^\dag \Omega)^2] = 1$ and the linear entanglement vanishes, $\mathcal{E}_i[\Omega^{\rm pure}]=0$, as must be the case.

\subsubsection{Entanglement generated through scattering}

We now extend our calculation to the quantity ${\cal E}$ generated through the S-matrix. Firstly, we consider four-scalar scattering through the following quartic contact term, 
\bea
\mathcal{L}_{\rm contact} = \frac{g^2}{M_X^2} c_{ij}^{kl} \phi_i \phi_j \phi_k \phi_l\,.
\label{eq:LEFT}
\eea
This can be generated in the UV through, for instance, the exchange of a heavy scalar $X$. 
The initial state~\eqref{eq:omegaapp} and initial state entanglement~\eqref{eq:Ein} are given above.  We are now interested in the final state density matrix and its corresponding ${\cal E}$ value, the definitions of which we repeat here for convenience:
\bea 
\hat{\rho}^f_{AB} &=& \hat{S} |\Omega \rangle \langle \Omega |\hat{S}^\dagger \,, \qquad\quad \mathcal{E}^f[\Omega] = 1 - \frac{1}{\Fn^2} \text{Tr}_A [\text{Tr}_B \hat{\rho}^f_{AB}]^2\,,\\[0.4em] 
\text{Tr}_B   \hat{\rho}_{AB}^f  &=& \frac{1}{\Fn} \sum_c \int_u \langle u, c|_B \hat{S} |\Omega \rangle \langle \Omega | \hat{S}^\dag |u,c \rangle_B\,.\nonumber
\label{eq:EFTrhof}
\eea
 Once again, the normalization $N$ is fixed by the initial state, see Eq.~\eqref{eq:Ein}. 
To evaluate Eq.~\eqref{eq:EFTrhof}, a useful trick is to insert the identity operator on Fock space, which we expand out to work at lowest order in perturbation theory: $\hat{S} |\Omega \rangle =\hat{\mathbbm{1}}_1 \hat{S} |\Omega \rangle + \hat{\mathbbm{1}}_2 \hat{S} | \Omega \rangle  + \cdots$, where $\hat{\mathbbm{1}}_1$ denotes the identity on the subspace of one-particle states, $\hat{\mathbbm{1}}_2$ on two-particle states, etc. 
At present, the $\hat{\mathbbm{1}}_1$ term does not contribute and the leading-order sum over states becomes (where the $q$ denote final momenta and $i,j$ are internal flavor indices): 
\be 
\begin{aligned}
\mathbbm{1}_2 \hat{S} |\Omega \rangle &= \sum_{ijab} \int_{q_1} \int_{q_2}  | q_1,i; q_2, j \rangle \langle q_1,i;q_2,j | (1+i \hat{T}) \Omega_{ab} |k_1,a ; k_2 ,b \rangle \\ 
& = |\Omega \rangle +  2\pi i \sum_{ijab} \int_{q_2} \frac{1}{2E_{q_1} }   \Omega_{ab}  \delta (E_{k_1}+E_{k_2} - E_{q_1} - E_{q_2} ) \\
&\qquad\qquad\qquad\qquad\qquad\qquad\times\mathcal{A}_{ij}^{ab} (k_1 k_2 \rightarrow q_1 q_2)  |q_1,i;q_2,j \rangle  \Big|_{q_1 = q_2 - k_1 - k_2}\,,
\end{aligned}
\ee

\noindent and similarly for the conjugate $\langle \Omega | \hat{S}^\dagger \mathbbm{1}_2 $. 
Plugging into 
Eq.~\eqref{eq:EFTrhof}, we compute the reduced density matrix:
\begin{equation}
\hspace{-1mm}
\begin{aligned}
N \hat{\rho}_A =  \sum_{jaa'} &\int \frac{d^3u}{ 2 E_u}  (2 E_{k_2})^2    \Omega_{ac} \Omega_{a'j}^\dag  (2\pi)^3  (\delta^3 (k_2 - u))^2 |k_1,a \rangle_A \langle k_1, a' |_A \\ 
&  {-}   \Bigg{[}i \pi \!\! \sum_{iaa'bj} \!\int\!\! \frac{d^3u}{2 E_u E_{q_1}}  2 E_{k_2}  \delta^3 (k_2{-} u)    \Omega_{a'j}\Omega_{ab}^\dag  \delta(E_{k_1}{+}E_{k_2} {-} E_{q_1} {-} E_u) \\
&\qquad\qquad \times \mathcal{A}_{ij }^{ab \dag} (q_1 u {\rightarrow} k_1 k_2 )|k_1,a' \rangle_A \langle q_1, i |_A {+}  \text{h.c.}\Bigg{]}_{q_1 {=} k_1 {+}k_2{-}u}\\ 
&+  \Bigg{[}\sum_{iaa'bb'j} \int \frac{d^3u \, \pi^2 \Omega_{ab}\Omega_{a'b'}^\dag}{(2\pi)^3 2 E_u E_{q_1}^2}   (\delta (E_{k_1} {+} E_{k_2} {-} E_{q_1} {-} E_{u}))^2 \\
&\qquad\qquad \times\mathcal{A}^{ab}_{ij} (k_1 k_2 {\rightarrow} q_1 u)  \mathcal{A}_{ij }^{a'b' \dag} (q_1 u {\rightarrow} k_1 k_2 )  | q_1,i \rangle_A \langle q_1, i |_A \Bigg{]}_{q_1 {=k_1 {+}k_2{-}u}}\,.
\label{eq:Nrhoint}
\end{aligned}\hspace{-2mm}
\end{equation}

\noindent Critically, note that, at tree level, the fourth term is $O(g^4)$ as it arises from the product of two \emph{four-point} amplitudes, while the second and third terms are $O(g^2)$. Computing the momentum integrals, we have to leading order,
\begin{equation}
\begin{aligned}
\Fn \, \hat{\rho}_A =\;&  2 E_{k_2}  V \sum_{jaa'}    \Omega_{aj} \Omega_{a'j}^\dag   |k_1,a \rangle_A \langle k_1, a' |_A \\ 
&-   2\left[ i\frac{T}{2  E_{k_1}}   \sum_{iaa'bj}    \Omega_{a'j}\Omega_{ab}^\dag   \mathcal{A}_{ij }^{ab \dag} (k_1 k_{2} \rightarrow k_1 k_2 )|k_1,a' \rangle_A \langle k_1, i |_A  + \text{h.c.}\right] + O(g^4),
\label{eq:rhoA}
\end{aligned}
\end{equation}
where we have used $(2\pi)^3  \delta^3 (0) \equiv V$ in the first line. We can now understand, pragmatically, how the kinematic limit arises; integrating over $u$ momentum in Eq.~\eqref{eq:Nrhoint} selects $u=k_2$, so that $q_1 = k_1+k_2 -u = k_1$. The energy delta function thus becomes $\delta(0)$ and we define $2\pi  \delta (0) \equiv  T$. The result in Eq.~\eqref{eq:rhoA} may now be squared and plugged into Eq.~\eqref{eq:EFTrhof} to compute {${\cal E}$} to leading order $O(g^2)$, which after   
evaluating the integrals and summing of internal symmetry indices, becomes
\begin{equation}
\begin{aligned}
\Fn^2 \, \text{Tr}_A \left[ \hat{\rho}_A^2 \right] 
=\;& (2 E_{k_1}  2 E_{k_2} V^2 )^2 \, \text{Tr} [(\Omega^\dag \cdot \Omega)^2] \\[0.3em]
&- 4 \, (2 E_{k_2}  2  E_{k_1} V^3 T)\,  \text{Im}\, \Bigl[ \sum_{abij}  \Omega_{ab}  \mathcal{A}_{ij }^{ab } (k_1 k_{2} \rightarrow k_1 k_2 ) (\Omega^\dag \cdot \Omega  \cdot \Omega^\dag)_{ij}  \Bigr] + O(g^4) \,,
\end{aligned}
\label{eq:N-rho-final}
\end{equation}
using $i (X-X^\dag) = -2 \, \text{Im} X$.  The resulting change in {$\cal E$} generated by scattering to leading order becomes Eq.~\eqref{eq:DeltaE}, namely,
\begin{equation}
\Delta \mathcal{E}_{\rm contact} [\Omega] =4\left(\frac{1}{2 E_{k_1} 2 E_{k_2}} \frac{T}{V}  \right)  \, \text{Im} \Bigl[ \sum_{abij} \Omega_{ab}  \mathcal{A}_{ij }^{ab } (k_1 k_{2} \rightarrow k_1 k_2 ) (\Omega^\dag \cdot \Omega  \cdot \Omega^\dag)_{ij} \Bigr]+ O(g^4)\,,
\label{eq:deltaE-contact}
\end{equation}
where we have subtracted off the the initial state entanglement fond in Eq.~\eqref{eq:Elinscat}, i.e., $\mathcal{E}_i = 1 - \text{Tr} \left[ (\Omega^\dag \Omega)^2 \right]$. 
For a pure state $\Delta \mathcal{E}_{\rm contact} [\Omega^{\rm pure}] \propto \text{Im}  \mathcal{A}  \propto \text{Im}  \left(g^2 c / M_X^2 \right) = 0$ since the coupling is real, and the leading-order contribution arises at $O(g^4)$, i.e., at one-loop order. However $\mathcal{E}_{\rm contact} [\Omega]$ will generally be nonvanishing for an impure initial state:
\bea
\Delta \mathcal{E}_{\rm contact} [\Omega ] =4  \left(\frac{1}{2 E_{k_1} 2 E_{k_2}} \frac{T}{V}  \right)  \,  \frac{g^2}{M_X^2} \text{Im} \Bigl[ \sum_{abij} \Omega_{ab}  c^{ab}_{ij}  (\Omega^\dag \cdot \Omega \cdot \Omega^\dag)_{ij}  \Bigr]+ O(g^4)\,,
\eea
where we have plugged in the amplitude from Eq.~\eqref{eq:LEFT}. 
We could of course dress the contact term with derivatives, in which case we can view this result in an EFT context: at linear order in the Wilson coefficients, the contribution to ${\cal E}$ generated by an EFT vanishes for a pure state, for scattering below the mass scale of the ultraviolet completion.

The $O(g^4)$ contribution to ${\cal E}$ includes both the one-loop part of the amplitude ${\cal A}^{1\text{-loop}}$, as well as terms going like the square of the tree-level amplitude,
\begin{equation}
\begin{aligned}
\label{eq:E-final}
\Delta \mathcal{E}[\Omega]_{O(g^4)} &= {4}\,{\mathcal{N}}\sum_{ijmn} \mathrm{Im} \big{[} \Omega_{ij} \,\mathcal{A}^{ij \, 1\text{-loop} }_{mn} \, (\Omega^\dagger \cdot \Omega \cdot \Omega^\dagger)_{nm}  \big{]}\\
&-{2}\,{\mathcal{N}^2}\, \sum_{ijmn}\sum_{i^\prime j^\prime m^\prime n^\prime}\bigg\{ \delta_{m m^\prime} \,\Omega_{ij}(\Omega^\dagger)_{j^\prime i^\prime}\,\mathcal{A}^{ij \, \rm{tree}}_{m n} \,(\Omega^\dagger \cdot \Omega)_{nn^\prime}\mathcal{A}^{\dagger\,i^\prime j^\prime \, \rm{tree}}_{m^\prime n^\prime}\,
\\ &\qquad\qquad\qquad\qquad\quad +\delta_{n n^\prime}\,\Omega_{ij}(\Omega^\dagger)_{j^\prime i^\prime}(\Omega\cdot \Omega^\dagger)_{m^\prime m}\,\mathcal{A}^{ij  \, \rm{tree}}_{mn}\mathcal{A}^{\dagger\,i^\prime j^\prime\,  \rm{tree}  }_{m^\prime n^\prime}\\
&\qquad\qquad\qquad\qquad\quad -\mathrm{Re}\Big[{}\Omega_{ij}\Omega_{i^\prime j^\prime} (\Omega^\dagger)_{n^\prime m}\, \mathcal{A}^{ij\, \rm{tree}}_{mn} \, (\Omega^\dagger)_{nm^\prime}\mathcal{A}^{i^\prime j^\prime \, \rm{tree}}_{m^\prime n^\prime}  \,\Big{]}  \bigg\} \,,
\end{aligned}
\end{equation}
where all amplitudes are in the forward limit {and we recall that $\mathcal{N}$ is defined in Eq.~\eqref{eq:calN}.}

\subsection{Wave Function Derivation}
\label{sec:wave-function-derivation}

In the previous derivation, we have encountered the factor of $\mathcal{N} = \left(\frac{1}{2 E_{k_1} 2 E_{k_2}} \frac{T}{V}  \right)$, which contains the spacetime volume. This factor comes from the scattering plane wave states, which have infinite norm.  
It is replaced with a finite normalization in the case of wave packet states (see, e.g., Ref.~\cite{Low:2024hvn}), which we now consider. That is, we do not scatter purely plane waves but states $|\psi\rangle$ with a given flavor index $a$,
\begin{align}
\label{eq:WavefunctionsDefs}
	|\psi, a\rangle \equiv \int_p \psi(p)|p,a\rangle, 
	\quad 
	\int_p \equiv\int \hat{d}^4p\,\, \hat{\delta}(p^2-m^2)\Theta(p^0),
\end{align}
where the hatted notation means $\hat{d}x= dx/2\pi$ and $\hat{\delta}(x) = 2\pi\delta(x)$, so that we do not need to carry around factors of $(2\pi)$~\cite{Kosower:2018adc}. 
The normalization is such that
\begin{align}
	\langle \psi,b|\psi,a\rangle = \delta^{b}_a \int_p ||\psi(p)||^2 =  \delta^{b}_a\,,
 \end{align}
It is also clear that from the definitions that the momentum-space representation of the wave function is $\langle k,b|\psi,a\rangle = \psi(k)\delta^{b}_a $. The most general pure initial state is given by
\begin{align}
	|{\rin}\rangle = \sum_{ab}\int_{k_1,k_2} \psi_{12}(k_1,k_2)\Omega_{ab} |k_1,a;k_2,b\rangle\,,
\end{align}
which in principle could be entangled in momentum and flavor space. We want start with a product state in momentum space $\psi_{12}(k_1,k_2) = \psi_{1}(k_1)\psi_2(k_2)$ while still permitting entanglement in flavor space,
\begin{equation}
	|{\rin}\rangle \Rightarrow |\Omega\rangle \equiv  \sum_{ab}\int_{k_1,k_2} \psi_{1}(k_1)\psi_2(k_2)\Omega_{ab} |k_1,a;k_2,b\rangle = \sum_{ab} \Omega_{ab} |\psi_1,a;\psi_2,b\rangle\,.
\end{equation}
The normalization of these states $\langle \Omega | \Omega \rangle = 1$ implies that $\sum_{ab} \Omega^\dagger_{ab} \Omega_{ab} =1$.  We then evolve this initial state via the S-matrix $|{\rm  out}\rangle = \hat S |{\rm in}\rangle = (1+i\hat T) |{\rm in}\rangle$ and keep terms up to linear order in $\hat T$. Performing the partial trace over $\mathcal{H}_B$,  we obtain the following reduced density matrix,
\be 
\begin{aligned}
\hspace{-0.5mm}	\rho_{A} =\;& \sum_{crs} \Omega_{rc}\Omega^\dagger_{sc} |\psi_1,r\rangle \langle \psi_1,s|\\
	&\! +  \Bigg{[}i\, \sum_{abi}\sum_{nm} \Omega_{ab} \Omega^\dagger_{nm}
	\!\!\!\!\!\!\!\int\limits_{q_1,q_2,k_1,k_2} 
  \!\!\!\!\!\!\!  \hat{\delta}^{(4)}(k_1{+}k_2{-}q_1{-}q_2)\\
  &\qquad\qquad\quad\times\psi_1(k_1)\psi_2(k_2)\psi^\dagger_2(q_2)\cA^{ab}_{im}(k_1{+}k_2{\rightarrow}q_1{+}q_2) |q_1,i\rangle\langle \psi_1,n| + \text{h.c.}\Bigg{]}
\end{aligned}
\ee
For the trace of $\rho_A^2$, it is also useful to define the smeared amplitude, which will be an object acting purely on the flavor space,
\begin{align}
	{\bm \cA}^{ab}_{cd} \,\,\,\equiv\!\!\!\!\!\!
    \int\limits_{p_1,p_2,p_3,p_4} \hat{\delta}^{(4)}(p_1{+}p_2{-}p_3{-}p_4) \psi_1(p_1)\psi_2(p_2)\psi_1^\dagger(p_3)\psi_2^\dagger(p_4)\cA^{ab}_{cd}(p_1{+}p_2{\rightarrow}p_3{+}p_4)\,.
\end{align}
Note that this object has only initial wave packets, which are evaluated on the initial and final momenta.
Squaring the reduced density matrix and finally tracing over $\mathcal{H}_A$ leads to 
\begin{equation}
	\text{tr}[\rho_{A}^2] = \sum_{crs}\Omega_{rc}\Omega^\dagger_{sc} 
	\left[
	 \sum_{c'r's'}\Omega_{sc'}\Omega^\dagger_{rc'}  +  2i\sum_{abi}\sum_{nm} \Omega_{ab} \Omega^\dagger_{nm}
	\left(	{\bm \cA}^{ab}_{im} \delta^{s}_i \delta^n_r - 
	{\bm \cA}^{ib}_{nm}\delta^s_a\delta^i_r \right)\right]\,,
\end{equation}	
which can be put in simplified notation, $\Omega_{ab} \equiv |{\bm \Omega}\rangle $ and $\Omega^\dagger_{nm} \equiv \langle {\bm \Omega}|$,
to obtain 
\begin{align}
\mathcal{E}[{\bm \Omega}] = 1 - \langle {\bm \Omega} | {\bm \Omega} {\bm \Omega}^\dagger |{\bm \Omega}\rangle  + 4\,\text{Im}[\langle {\bm \Omega} |  {\bm \Omega} {\bm \Omega}^\dagger {\bm \cA}  | {\bm \Omega}\rangle] + O(\cA^2)\,.
\end{align}
In the case where the initial state is a product state, $(\Omega\cdot \Omega^\dagger \cdot \Omega)_{rc'} = \Omega_{rc'}$, i.e., ${\bm \Omega} {\bm \Omega}^\dagger |{\bm \Omega}\rangle= |{\bm \Omega}\rangle$. Then
\begin{align}
	\mathcal{E}[{\bm \Omega}] = 4\,\text{Im}[\langle {\bm \Omega} | {\bm \cA} | {\bm \Omega}\rangle] + O(\cA^2)\,.\label{eq:wavepacket}
\end{align}
{Comparing with Eq.~\eqref{eq:DeltaEPure}, we see that in the wave packet regularization, ${\cal N}$ has been absorbed into the definition of the smeared amplitude $\boldsymbol{\cal A}$; that is, $\boldsymbol{\cal A}$ contains the normalization of the wave packets, which in the plane-wave limit becomes the ordinary amplitude times ${\cal N}$. See Refs.~\cite{Low:2024hvn,Kowalska:2024kbs,Kowalska:2025qmf,Peschanski:2026edo} for more discussion on the meaning of the wave packet normalization factor.}
Coming back to $\text{tr}[\rho_{A}^2]$, if we were to calculate the quantum Tsallis entropy, we would need the $n$th power instead,
\bea
\text{tr}[\rho_{A}^n] = \langle {\bm \Omega} | ({\bm \Omega} {\bm \Omega}^\dagger)^{n-1} |{\bm \Omega}\rangle + 2n\,\text{Im}[\langle {\bm \Omega} |  ({\bm \Omega} {\bm \Omega}^\dagger)^{n-1} {\bm \cA}  | {\bm \Omega}\rangle]\,,
\eea
which in the case of a product initial state reduces to $\mathcal{E}_n[{\bm \Omega}] = \frac{2n}{n-1}\,\text{Im}[\langle {\bm \Omega} | {\bm \cA} | {\bm \Omega}\rangle]$.

The details of the wave function in the smeared amplitude should not be relevant to our discussion. However, to recover the usual story when scattering infinite-norm states, one needs to choose a wave function $\psi_1 (p) \propto \hat{\delta}^{(3)}(p-P_\rma)$ where $P_\rma$ is the plane-wave momentum. This will take the initial $|\rin\rangle$ to have the normalization as in Eq.~(\ref{eq:Ein}). Alternatively, one could also choose $\psi_1 (p) = |\psi_1 (p)| e^{ib\cdot p}$ with an impact parameter $b$ such that $|\psi_1 (p)|^2$ is a Gaussian localized around $P_\rma$ similar to Ref.~\cite{Kosower:2018adc}. 
In this case, the smeared amplitude ${\bm \cA}^{ab}_{cd}$ becomes an integral over the transferred momentum $q$, and in the limit of small $q$, it becomes an eikonal amplitude in impact-parameter space. Another option is to use square wave packets with size $L$ ($1/\delta_p$) in the transverse (longitudinal) direction as shown in Ref.~\cite{Low:2024hvn}, which recovers the plane-wave limit when $\delta_pL \gg 1$ and $\delta_p \to 0$.
\vspace{\baselineskip}
\section{Amplitudes' Positivity and Entanglement} Using our calculations from the previous section, we are now poised to demonstrate the primary result of this paper summarized in Eq.~\eqref{eq:goal},
i.e., that positivity of ${\cal E}$ is related to positivity of elastic amplitudes.  While one direction of this relation is immediate---in a unitary theory, both the optical theorem and positivity of entropy hold---the converse will interest us here, in that demanding that ${\cal E}$ is positive, without explicitly assuming unitarity or that it is a well defined entropy, is sufficient for recovering positivity of ${\rm Im}\,{\cal A}$.

We are interested in contributions to ${\cal E}$ that are generated by the S-matrix, so as such we will consider a state where the two initial particles are unentangled, so that ${\cal E}_i =0$. 
By Schmidt decomposition, this means that there exist unique vectors $\ket{\alpha}_A\in{\cal H}_A$ and $\ket{\beta}_B \in {\cal H}_B$---given by some sums over the basis vectors $\ket{a}_A$ and $\ket{b}_B$, respectively---such that $\ket{\Omega^{\rm prod}} = \ket{k_1,\alpha}_A \otimes \ket{k_2,\beta}_B$ is a direct product state. 
In that case, $(\Omega\cdot\Omega^\dagger\cdot\Omega)_{ab}$ becomes $\Omega_{ab}$, so from Eq.~\eqref{eq:DeltaE} the leading-order change in $\cal E$ generated by scattering is  
\be 
\begin{aligned}
\Delta \mathcal{E}[\Omega^{\rm prod} ] &= 4\, {\cal N}  \,  \text{Im} \,\mathcal{A}^{\alpha \beta}_{\alpha\beta} (k_1 k_2 \rightarrow k_1 k_2 )\,.
\end{aligned}
\label{eq:DeltaEPure}
\ee 
The amplitude on the right-hand side is kinematically forward, as well as elastic in the internal quantum numbers.
We now assume positivity of $\cal E$ as an axiom, in which case we find critically that $\Delta \mathcal{E}[\Omega^{\rm prod} ] \geq 0 $ since $\mathcal{E}_i = 0$ and $\mathcal{E}_f \geq 0$. Thus, positivity of the shift in ${\cal E}$ also implies positivity of the imaginary part of the forward amplitude (which need not have been positive a priori without unitarity, $\hat{S}^\dag \hat{S} = \hat{\mathbbm{1}}$). 
If we were to invoke the optical theorem at this point, we would have $\text{Im} \,\mathcal{A}^{\alpha\beta}_{\alpha\beta} (k_1 k_2 \rightarrow k_1 k_2 ) = 2E_{\rm cm}p_{\rm cm}\sigma_{\alpha\beta}$,
where $\sigma_{\alpha\beta}$ is the cross section for $|\alpha \rangle_A\otimes |\beta \rangle_B\rightarrow{\rm anything}$ and which as a physical area is by definition positive, $E_{\rm cm}$ is the energy of the two-particle initial state in the center-of-mass frame, and $p_{\rm cm} = v_1 E_{k_1} = v_2 E_{k_2}$ is the momentum of either of the initial particles.
Thus, the optical theorem implies positivity of the entanglement for elastic scattering,
\begin{align}
\Delta {\cal E}[\Omega^{\rm prod}] = 2(T/V)(v_1+v_2)\sigma_{\alpha\beta} \geq 0\,.
\end{align}
{See App.~\ref{app:norm} for a discussion of the normalization under the alternative choice of fixing it using the final state. Such projective measurement choices have been made in other papers on entanglement in scattering, with the result being a replacement of the total cross section with either the elastic~\cite{Low:2024hvn} or inelastic~\cite{Kowalska:2024kbs,Kowalska:2025qmf} cross section. However, conventions consistent with ours, Ref.~\cite{Cheung:2023hkq} similarly found that forward momentum kinematics, and the total cross section among internal quantum numbers, was selected by the entanglement calculation (see, e.g., their footnote [23]).}
Note that for this product initial states, this leading-order tree-level entanglement is only nonvanishing when it has support on the resonance (as shown in App.~\ref{app:B}). 
However, this conclusion generalizes to all weakly coupled theories, as one can view any loop-level process as an infinite sum over trees, as shown in App.~B of Ref.~\cite{Arkani-Hamed:2020blm}.  
We leave a detailed study of higher-order contributions to future work.

These conclusions generalize to the quantum Tsallis entropy ${\cal E}_n = \frac{1}{n-1}(1-{\rm Tr}_A[\hat \rho_A^n])$, for which the linearized entanglement is the $n\,{=}\,2$ case. 
As above, ${\cal E}_n$ is nonnegative and vanishes for a pure state.
For an initial product state, as shown in the Appendix, we have $\Delta {\cal E}_n[\Omega^{\rm prod}] = \frac{2n}{n-1}{\cal N} \,\text{Im} \,\mathcal{A}^{\alpha \beta}_{\alpha\beta} (k_1 k_2 \rightarrow k_1 k_2 )$ for all integers $n\geq 2$.

Thus, we have shown that positivity of ${\cal E}$---which in a unitary theory corresponds to the linear entropy---and positivity of the amplitudes are {\it equivalent}, as advertised in Eq.~\eqref{eq:goal}.
Both are manifestations of unitarity, but in a priori very different contexts, namely dispersion relations and consistency of quantum information. 
From this perspective, all positivity bounds on EFTs---including their many consequences including applications to quantum gravity, the EFT-hedron~\cite{Arkani-Hamed:2020blm}, etc.---are all quantum information theoretic statements, demanding consistency of the entanglement produced among the decay products.

\vspace{\baselineskip}
\section{Disentanglers}For entangled initial states, $\Delta \mathcal{E}$ in Eq.~\eqref{eq:DeltaE} need not always be positive. Such ``disentangler states'' occasionally arise when the S-matrix itself acts as a  ``Maxwell's demon'' or ``disentangler,'' decreasing the entanglement of the final state relative to the initial one. Similar states were explored in Ref.~\cite{Cheung:2020uts}, and their existence can be anticipated from the fact that in any quantum theory the action of the S-matrix on a state may be reversed through acting with $\hat{S}^\dag$; such objects have found extensive use in tensor network contexts~\cite{Evenbly:2007hxg,Bao:2015uaa}.

We now demonstrate the existence of such disentangler states and highlight their role in hindering us from deriving the {\it generalized}  optical theorem from entanglement~\footnote{Understanding the connection between entanglement and the generalized optical theorem is particularly interesting in light of recent positivity bounds that make use of it~\cite{Trott:2020ebl,Zhang:2020jyn,Arkani-Hamed:2021ajd,Freytsis:2022aho}.}. %
Recall that the generalized optical theorem is the
statement that ${\cal A}_{ab}^{ij}$ is a positive definite matrix,
a property that is preserved under unitary transformations. 

Let us Schmidt-decompose the initial state,
$|\Omega\rangle=\sum_{k}\lambda_{k}|\alpha_{k}\rangle\otimes|\beta_{k}\rangle$, i.e., $\Omega_{ij}=\sum_{k}\lambda_{k}\alpha_{i}^{(k)}\beta_{j}^{(k)}$,
for some real $\lambda_{k}\geq0$ and orthonormal vectors $|\alpha_k\rangle$ and
$|\beta_k\rangle$, 
so Eq.~\eqref{eq:DeltaE} now becomes
\begin{equation}
\hspace{-1mm}\Delta{\cal E}[\Omega]\,{=}\,4 \, {\cal N} \,{\rm Im}\bigg[\sum_{kk'}\lambda_{k}\alpha_{j}^{*(k)}\beta_{i}^{*(k)}\!{\cal A}_{ab}^{ij}\lambda_{k'}^{3}\alpha_{a}^{(k')}\beta_{b}^{(k')}\bigg].\hspace{-1mm}
\label{eq:DeltaEschmidt}
\end{equation}
Note that without internal flavor indices, we would have $\Delta{\cal E} \propto \text{Im} \mathcal{A}(k_1 k_2 \rightarrow k_1 k_2) \propto \sigma \geq 0$, in which case disentangler states would not arise.
However, in the presence of the internal degrees of freedom considered here, this is no longer the case
as we will now demonstrate.

Let us write the basis of $|i\rangle \otimes |j\rangle$ states as $|I\rangle$, each denoted by some vector $v_{I}$
\be
\Delta{\cal E}[\Omega]=4\,{\cal N} \,{\rm Im}\bigg[\sum_{k,k'}\lambda_{k}\lambda_{k'}^{3}v_{I}^{(k)*}{\cal A}_{I'}^{I}v_{I'}^{(k')}\bigg]\,.
\ee
Without loss of generality, we take each  $v_{I}$
to be real (a complex $v_{I}$ can made real via a unitary transformation, which we absorb into ${\cal A}$). 
Thus, we can pull everything but the amplitude out of the imaginary part of Eq.~\eqref{eq:DeltaEschmidt}, yielding
\begin{equation}
\Delta{\cal E}[\Omega]=4\,{\cal N}\, \sum_{k,k'}\lambda_{k}\lambda_{k'}^{3}x^{(k)}\cdot x^{(k')}\,, 
\end{equation}
where 
we have used the generalized optical theorem to decompose ${\rm Im}\,{\cal A}_{I'}^{I}=M^{IJ}M_{JI'}$
for some matrices $M$ and defined the vectors $x^{(k)}=M\cdot v^{(k)}$. 
The $M$ matrices---which can be taken to be real~\cite{Zhang:2020jyn,Arkani-Hamed:2021ajd}---describe the cuts of the four-particle amplitudes, that is, the (real and imaginary parts of) amplitudes for a given two-particle state described by an element of the Schmidt basis to go to some exchanged state.
By simple algebra and relabeling of the indices $k,k'$, we have:
\begin{equation}
\Delta{\cal E}[\Omega]\,{=}\,\frac{\cal N}{4}\sum_{k,k'}\left[(\lambda_{k}{+}\lambda_{k'})^{4}{-}(\lambda_{k}{-}\lambda_{k'})^{4}\right]x^{(k)}\cdot x^{(k')}.
\label{eq:SchmitDeltaE}
\end{equation}
Since the generalized optical theorem is both necessary
and sufficient for a healthy perturbative ultraviolet completion, it is possible to design
a completion with arbitrary $x^{(k)}\cdot x^{(k')}$. One can think
of $x^{(k)}\cdot x^{(k')}$ as defining a Riemannian metric $g_{kk'}$, i.e., $g_{kk'}$ is  positive definite.
For an \emph{un}-entangled initial state, the
Schmidt decomposition has only one nonzero value of $\lambda$, so that $\Delta{\cal E}$
is manifestly nonnegative since all the $\lambda_{k}-\lambda_{k'}$
differences vanish. However, for general $|\Omega \rangle$, we now demonstrate that disentangler states for which
$\Delta{\cal E}[\Omega] < 0$ can arise. 

\begin{figure}
\begin{center}
\includegraphics[width=6.5cm]{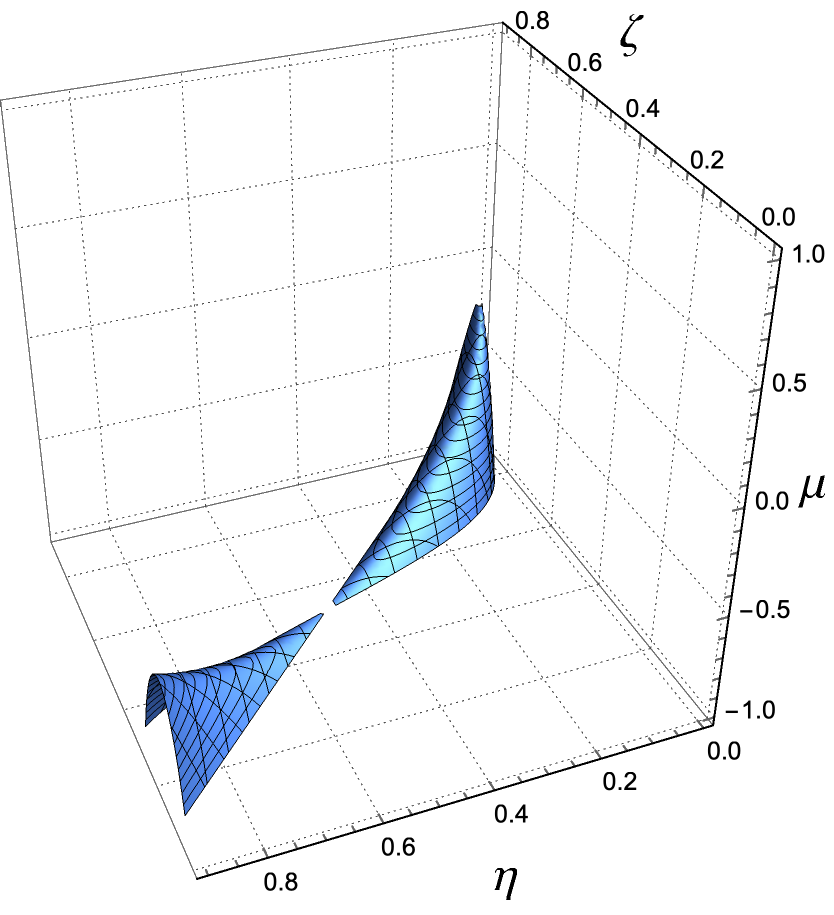}
\caption{\label{fig:Maxwell} The blue surface and volume below it corresponds to the parameter space described in text for a disentangler state, for which $\Delta{\cal E}<0$. }
\end{center}
\end{figure}

As an illustration that $\Delta{\cal E}$ can be negative, consider for simplicity 
a scenario with only two nonzero $\lambda_{k}$ values in the Schmidt
basis. For example, each particle could be a qubit, leaving us with a four-dimensional joint Hilbert space, and
we can choose to scatter an incoming Bell state, e.g., $(|1\rangle|1\rangle{+}|2\rangle|2\rangle)/\sqrt{2}$
or $(|1\rangle|2\rangle{-}|2\rangle|1\rangle)/\sqrt{2}$.
Let us define the ratio $\lambda_{2}/\lambda_{1} = \zeta/(1-\zeta) > 0$ ($\zeta \in [0,1]$), and parameterize the vectors $x^{(1)}$ and $x^{(2)}$ with $\mu\in [-1,1]$ the cosine of the angle between them and $|x^{(2)}|/|x^{(1)}| = \eta/(1-\eta)$ the ratio of lengths (for $\eta \in [0,1]$).
Note that $\zeta = 1$ or $0$ corresponds to a pure state and $\zeta = 1/2$ to maximally mixing. We indeed find parameters leading to disentanglers, as shown by the region on and under the blue surface in Fig.~\ref{fig:Maxwell}. Note that disentanglers only appear when $\mu < 0$, corresponding to the physical picture where vectors $x^1$ and $x^2$ are pointing away from each other such that the element $g_{12} <0$ is possible. 

We observe in passing that if $x^{(k)}\cdot x^{(k')}$ is diagonal, equal to $\gamma_{k}\delta_{kk'}$ for $\gamma_{k}\,{>}\,0$, then from Eq.~\eqref{eq:SchmitDeltaE} $\Delta{\cal E}[\Omega]\,{=}\,4{\cal N}\sum_{k}\gamma_{k}\lambda_{k}^{4}\,{\geq}\,0$ for any $\lambda_k$.
This situation corresponds to the case where there is no mixing among the exchanged states to which each two-particle state in the Schmidt basis couples, that is, the scattering describes a sum over disjoint superselection sectors.

Moreover, for projector states~\footnote{The result about projector states and Maxwell's demon was also found in Ref.~\cite{Cheung:2023hkq} in the context of a product initial two-particle state that is not necessarily pure.}, where $\lambda_k \,{=}\, c$ for $k\,{\leq}\, n$ for some $n$ and $\lambda_{k}\,{=}\,0$ for $k\,{>}\,n$, $\Delta {\cal E}$ is positive for all S-matrices: $\Delta {\cal E}\,{=}\, 2{\cal N}c^4 |X^{(n)}|^2\,{\geq}\, 0$, defining the vector $X^{(n)}\,{\equiv}\, \sum_{k=1}^n x^{(k)}$. 
The converse is also true: if $\lambda_k$ does not take the projector form, it is a disentangler state for some S-matrix.
For $\lambda_k$ not a projector, there exist $c_1 \,{>}\, c_2$ for which we can write $\lambda_k \,{=}\, c_1$ for $k\;{\leq}\; n$ for some $n$ and $\lambda_{k+1}\,{=}\, c_2$. 
We pick the S-matrix such that $x^{(k)}\,{=}\, 0$ for $k\,{>}\,n\,{+}\,1$ and 
$X^{(n)}\,{\cdot}\, x^{(n+1)}\,{=}\,{-}\sqrt{|X^{(n)}|^2 |x^{(n+1)}|^2}$, as permitted by Cauchy's inequality. Defining 
$q\,{=}\,\sqrt{|x^{(n+1)}|^2/|X^{(n)}|^2}$, we have $\Delta{\cal E}/|X^{(n)}|^2\,{=}\,2(c_1\,{-}\,qc_2)(c_1^{3}\,{-}\,qc_2^{3})$.
Since $c_1>c_2>0$, it is always possible to choose $q$ such that $c_1/c_2 < q < c_1^3/c_2^3$, in which case $\Delta{\cal E}<0$. 
This construction completes the proof that the projector form for the Schmidt coefficients is both necessary and sufficient to prevent disentangler states---for which scattering acts as a Maxwell's demon---for arbitrary S-matrices.
 
\vspace{\baselineskip}
\section{Discussion and Future Directions}In this work we have shown that amplitudes' positivity---a consequence of unitarity in the form of the optical theorem---implies positivity of entanglement, and vice versa.
Specifically, we have shown that positivity of the shift in the quantity ${\cal E}$, which in a unitary theory would correspond to the linearized entropy, implies positivity of the imaginary part of the forward amplitude. Could there be other consequences of unitary that are related to entanglement? Could entanglement be used as a ``guiding principle'' of sorts for effective field theories?   This paper paves the way to such explorations. 

As discussed above, constraints on Wilson coefficients are traditionally derived from unitarity, which we can now relate to entanglement positivity.
Consider as a simple example collinear two-to-two photon scattering arising from the EFT operator $\frac{C}{M_\Phi^2} (F_{\mu \nu}F^{\mu \nu})^2$, which can be generated via a heavy scalar $\Phi$ dimensionfully coupled to $F_{\mu\nu}F^{\mu\nu}$. In Ref.~\cite{Adams:2006sv}, the positivity bound was shown to be $C>0$.
In the forward limit the photon has only transverse modes and behaves as a complex scalar. As shown in Sec.~\ref{sec:ent-2to2}, positivity of $
\Delta \mathcal{E}_{\rm \gamma \gamma} [\Omega^{\rm prod}] \propto C s^2 \delta (s-M_\Phi^2)$
implies the positivity bound ${\rm Im}\,{\cal A} > 0$, and indeed ${\rm Im}\,{\cal A} \propto C \delta(s-M_\Phi^2)$, where the IR EFT coefficient $C$ is identified with $g^2$ for UV coefficient $g$ controlling the coupling $\Phi\rightarrow\gamma\gamma$. We leave it to future work to extend such a study beyond this simplest of cases. For instance, one could consider a theory with two scalars, allowing for a CP phase, as well scattering through other operators considered in Ref.~\cite{Remmen:2019cyz}. Such studies could also lead to new insights for the emergence of symmetries, in a similar spirit as Refs.~\cite{Beane:2018oxh,Low:2021ufv}. It would also be interesting to explore more generally the entanglement structure of theories with decays, mass mixing, and particle-antiparticle oscillations.

\section*{Acknowledgments} 
\noindent 
We thank Cliff Cheung, Ian Low, and Zhewei Yin for comments.
This work was supported in part by the European Union's
Horizon research and innovation programme under the Marie Sklodowska-Curie grant agreements No.~860881-HIDDeN and No.~101086085-ASYMMETRY. R.A. was funded by the STFC grant ``Particle Physics at the Higgs Centre.''
The research of G.E. is supported by the National Science Foundation (NSF) Grant Number PHY-2210562, by a grant from University of Texas at Austin, and by a grant from the Simons Foundation. 
G.N.R. is supported by the James Arthur Postdoctoral Fellowship at New York University.
G.E. thanks the Aspen Center for Physics (supported by NSF grant PHY-2210452) for their hospitality while this work was in progress.

\section*{Data Availability Statement}

Data sharing is not applicable to this article as no datasets were generated or analyzed during the current study.

\appendix

{

\section{On the density-matrix normalization}
\label{app:norm}

In this Appendix, we briefly comment on how our results would have changed if we had required that the normalization factor $N$ in Eq.~\eqref{eq:entpower} is fixed for the \emph{final state}, instead of the initial state, as performed, e.g., in Ref.~\cite{Kowalska:2024kbs,Kowalska:2025qmf} and references therein. With these assumptions, Eq.~\eqref{eq:E-final} would still hold, with the only difference that $\mathcal{N}$ is no longer given by Eq.~\eqref{eq:calN} and depends on the S-matrix. Indeed, the equivalent expression would be, in this case,
\begin{align}
    N^\prime &=  \text{Tr}_A [\text{Tr}_B \hat{\rho}^f_{AB}]  = 2E_{k_2} 2 E_{k_1}V^2 -2 (2\pi)^4 \delta^{(4)}(0) \sum_{ijmn} \mathrm{Im} \big{[} \Omega_{ij} \,\mathcal{A}^{ij\,1\text{-loop}}_{mn} \, \Omega^\dagger_{mn}  \big{]}\\
        &+(2\pi)^4 \delta^{(4)}(0)\int_{q_1} \int_{q_2} (2\pi)^4 \delta^{(4)}(k_1+k_2-q_1-q_2)\sum_{ij}\sum_{i^\prime j^\prime}\Omega_{ij}\Omega_{i^\prime j^\prime}^\ast \mathcal{A}^{ij\,\text{tree}}_{mn} \,\mathcal{A}^{\dagger\,i^\prime j^\prime\,\text{tree}}_{mn} \,. \nonumber
\end{align}

\noindent where we use a prime to distinguish it from $N=2 E_{k_2} E_{k_1} V^2$, i.e., the normalization fixed for the initial-state density matrix. Therefore, the prefactor $\mathcal{N}$ should be replaced by $\mathcal{N}^\prime$, which is given up to $O(g^4)$ by
\begin{align}
\label{eq:Taylor-Nprime}
\mathcal{N}^\prime \equiv \dfrac{(2\pi)^4 \delta^{(4)}(0)}{N^\prime} = \mathcal{N} - 2 \sum_{ijmn} \mathrm{Im} \big{[} \Omega_{ij} \,\mathcal{A}^{ij\,1\text{-loop}}_{mn} \, \Omega^\dagger_{mn}  \big{]}+{O}(g^4)\,.
\end{align}

\noindent Recalling the expression for the entanglement entropy at this order,
\begin{align}
    \mathcal{E}_f^\prime[\Omega] = 1 - \bigg(\dfrac{\mathcal{N}}{\mathcal{N}^\prime} \bigg)^2 \mathrm{Tr}[(\Omega^\dagger\cdot \Omega)^2] + 4\mathcal{N}^\prime\, \text{Im} \Bigl[ \sum_{abij} \Omega_{ij}  \mathcal{A}^{ij\,1\text{-loop}}_{ab} (k_1 k_2 {\rightarrow} k_1 k_2 ) (\Omega^\dag \cdot \Omega \cdot \Omega^\dag)_{ab}  \Bigr]+{O}(g^4)\,,
\end{align}

\noindent where we keep the dependence on $\mathcal{N}^\prime$ explicit, as Eq.~\eqref{eq:Taylor-Nprime} implies additional terms of ${O}(g^4)$. If we now impose unitarity of the S-matrix, the optical theorem would then lead to a cancellation between the contributions to the second and third terms for a product initial-state, at this order in the expansion, since $\Omega^\dagger \cdot \Omega \cdot \Omega^\dagger = \Omega^\dagger$ in this case.  We leave further consideration of alternative normalization choices to other work~\cite{Low:2024hvn,Kowalska:2024kbs,Kowalska:2025qmf} and in the present paper adopt conventions consistent with Ref.~\cite{Cheung:2023hkq}, which yields results consistent with ours.
}

\section{Entanglement derivation for massive exchange}
\label{app:B}
In Sec.~\ref{sec:ent-2to2}, we computed the shift in ${\cal E}$ in two-to-two scalar scattering $\Delta \mathcal{E} [\Omega]$ that arises from a contact operator, namely Eq.~\eqref{eq:LEFT}, and have shown that this is equivalent to Eq.~\eqref{eq:DeltaE}. To verify the robustness of Eq.~\eqref{eq:DeltaE} we now compute $\Delta \mathcal{E} [\Omega]$ in the context of a theory with massive exchange, which reduces to Eq.~\eqref{eq:LEFT} in the low-energy limit:
\bea
\mathcal{L} = g f_{ijk} X_i \phi_j \phi_k\,.
\eea
Here $X$ is a heavy scalar, and the whole Hilbert space is assumed to be decomposed as $\mathcal{H} = \mathcal{H}_X \otimes \mathcal{H}_A \otimes \mathcal{H}_B$. 
Once again the initial state is given by 
\bea
|\Omega \rangle = \sum_{ij} \Omega_{ij} |\phi_i (k_1) \rangle_A \otimes |\phi_j (k_2) \rangle_B \equiv  \sum_{ij} \Omega_{ij}| \phi_i(k_1) \phi_j (k_2) \rangle\,,
\eea
This is equivalent to Eq.~\eqref{eq:Omegain}, but with a slight change of notation. We distinguish a state of the heavy scalar as $|X_m(Q) \rangle_X$. 
Since in the ultraviolet one can form on-shell $X$ particles as intermediate states, we must sum over $X$ when inserting the identity operator,
\bea
\hspace*{-2.em}\hat{\mathbbm{1}} = \sum_{m} \int_Q |X_m(Q) \rangle_X \langle X_m(Q)  |_X + \sum_{ij} \int_{q_1} \int_{q_2} |\phi_i (q_1) \phi_j (q_2) \rangle \langle \phi_i (q_1) \phi_j (q_2) | + \cdots\,,
\label{eq:identity}
\eea
which we write as $\hat{\mathbbm{1}} = \hat{\mathbbm{1}}_1 + \hat{\mathbbm{1}}_2$.
Since the $\hat{T}$ matrix will now generate $2\leftrightarrow 1$ processes involving $X$, 
we define amplitudes generically now as
\be 
\begin{aligned}
\langle p_{f}, a_f  |\hat{T}| p_{i},a_i  \rangle &= (2\pi)^4 \delta^4 \left(p_f - p_i \right) \cA^{a_i}_{a_f}\left( p_i \rightarrow p_f \right) \,,
 \end{aligned}
\ee
where $p_{f/i}$ and $a_{f/i}$ stand for any number of final and initial state momenta ($X$ or $\phi$) and internal symmetry indices. 
Inner products between states are $\langle \phi_i (p_1) |_A | \phi_j (p_2) \rangle_A = (2\pi)^3 \delta^3 (p_1-p_2) \delta_{ij}$ and $\langle X | \phi \rangle = 0$, since there is no $X$-$\phi$ mass mixing.
(Note that if we had $\mathcal{L} = m_{ij} \phi_i \phi_j + g f_{ijk} \phi_i \phi_j \phi_k$, such vector products would not vanish and one could study the entanglement generated in mass mixing and more generically in particle-antiparticle oscillations. Studying such systems using our framework could, e.g., shed new insights on entanglement generated in neutrino oscillations.)

We now compute the reduced density matrix by tracing out the $\phi$  in $\mathcal{H}_B$, $\hat{\rho}_A = \text{Tr}_B \hat{\rho}_{AB}$, as follows: 
\bea
\hat{\rho}_A = \frac{1}{\Fn}\sum_c \int_u \langle \phi_c (u)|_B (\hat{\mathbbm{1}}_1 \hat{S} +  \hat{\mathbbm{1}}_2 \hat{S}) |\Omega \rangle \langle \Omega | (\hat{S} \hat{\mathbbm{1}}^\dagger_1 + \hat{S}^\dagger \mathbbm{1}_2) |\phi_c (u)  \rangle_B\,.
\eea
The matrix elements $\langle u ,c |_B  \hat{\mathbbm{1}}_2 \hat{S} |\Omega \rangle$ and its conjugate are essentially equivalent to those of the previous contact-term computation (of course with the amplitude itself now as given by the Lagrangian in this subsection). 
We note that the contribution from the internal $X$ particle given by $\hat{\mathbbm{1}}_1$ vanishes,
\begin{equation}
\langle \phi_c(u) |_B   \hat{S} \hat{\mathbbm{1}}_1 |\Omega \rangle = 0 = \langle \phi_c(u) |_B  \hat{\mathbbm{1}}_1 \hat{S}  |\Omega \rangle\,,
\end{equation}
since in this theory there is no mass mixing or other way to convert $\phi \leftrightarrow X$.
This makes intuitive sense since because, \emph{to leading order}, neither the final nor the initial state has overlap with $\mathcal{H}_X$. Thus we conclude that the density matrix is as we previously computed,
\bea
\hspace*{-3.5em}&&\Delta \mathcal{E}_{\rm UV}[\Omega] =4 \left(\frac{1}{2 E_{k_1} 2 E_{k_2}} \frac{T}{V}  \right)  \, \text{Im} \Bigl[ \sum_{abij} \Omega_{ij}  \mathcal{A}^{ij}_{ab} (k_1 k_2 \rightarrow k_1 k_2 ) (\Omega^\dag \cdot \Omega \cdot \Omega^\dag)_{ab}  \Bigr]+ O(g^4)\,.
\eea
For a narrow $X$ resonance, the imaginary part of the amplitude has support on the delta function $\delta(s-M_X^2)$.

The toy model considered above has a realization as collinear two-to-two photon scattering. A Lagrangian with a dilaton coupling $\mathcal{L} \supset \frac{g}{f} \Phi F_{\mu \nu}F^{\mu \nu}$, permits $\Phi$-mediated light-by-light scattering, which at low energies generates the effective operator $\frac{C}{M_\Phi^2} (F_{\mu \nu}F^{\mu \nu})^2$, where $C=g^2/2f^2>0$; see, e.g., the model in Ref.~\cite{Remmen:2019cyz}. 
In the forward limit, all polarizations are transverse to all momenta and thus behave like a complex internal quantum number~\cite{Bellazzini:2016xrt}, so the computations above for a scalar theory can be applied. For product initial states with polarizations $\epsilon_1$ and $\epsilon_2$, the leading-order entanglement becomes
\be
 \Delta \mathcal{E}_{\rm \gamma \gamma} [\Omega^{prod}]\propto \frac{g^2 s^2}{2f^2} \, \text{Im} \Bigl[ \frac{  (\epsilon_1 \cdot \epsilon_2)(\epsilon_1 \cdot \epsilon_2)^\star + (\epsilon_1 \cdot \epsilon_2^\star)(\epsilon_2 \cdot \epsilon_1^\star) }{s-M_\Phi^2 + i \epsilon} 
\Bigr]  \propto C s^2 \delta (s-M_\Phi^2)\, ,
\label{eq:EEphoton}
\ee
where we have used the EFT matching to rewrite the coupling $g$ in terms of the Wilson coefficient $C$.



\bibliographystyle{utphys-modified}
\bibliography{entanglement}

@article{Peschanski:2026edo,
    author = "Peschanski, Robi and Seki, Shigenori",
    title = "{Entanglement in Elastic and Inelastic Two-particle Scatterings at High Energy}",
    eprint = "2601.22502",
    archivePrefix = "arXiv",
    primaryClass = "hep-th",
    month = "1",
    year = "2026"
}

@article{Low:2024hvn,
    author = "Low, Ian and Yin, Zhewei",
    title = "{Elastic cross section is entanglement entropy}",
    eprint = "2410.22414",
    archivePrefix = "arXiv",
    primaryClass = "hep-th",
    doi = "10.1103/PhysRevD.111.065027",
    journal = "Phys. Rev. D",
    volume = "111",
    pages = "065027",
    year = "2025"
}

@article{Kowalska:2024kbs,
    author = "Kowalska, Kamila and Sessolo, Enrico Maria",
    title = "{Entanglement in flavored scalar scattering}",
    eprint = "2404.13743",
    archivePrefix = "arXiv",
    primaryClass = "hep-ph",
    doi = "10.1007/JHEP07(2024)156",
    journal = "JHEP",
    volume = "07",
    pages = "156",
    year = "2024"
}

@article{Kowalska:2025qmf,
    author = "Kowalska, Kamila and Sessolo, Enrico Maria",
    title = "{Qubit entanglement from forward scattering}",
    eprint = "2510.04200",
    archivePrefix = "arXiv",
    primaryClass = "hep-ph",
    month = "10",
    year = "2025"
}

@article{Einstein:1935rr,
    author = "Einstein, Albert and Podolsky, Boris and Rosen, Nathan",
    title = "{Can quantum mechanical description of physical reality be considered complete?}",
    doi = "10.1103/PhysRev.47.777",
    journal = "Phys. Rev.",
    volume = "47",
    pages = "777",
    year = "1935"
}

@article{Bell:1964kc,
    author = "Bell, J. S.",
    title = "{On the Einstein-Podolsky-Rosen paradox}",
    reportNumber = "RX-1376",
    doi = "10.1103/PhysicsPhysiqueFizika.1.195",
    journal = "Physics Physique Fizika",
    volume = "1",
    pages = "195",
    year = "1964"
}

@article{CHSH_1969,
	title     = {Proposed Experiment to Test Local Hidden-Variable Theories},
	author    = {Clauser, John F. and Horne, Michael A. and Shimony, Abner and Holt, Richard A.},
	journal   = {Phys. Rev. Lett.},
	volume    = {23},
	issue     = {15},
	pages     = {880},
	numpages  = {0},
	year      = {1969},
	publisher = {American Physical Society},
	doi       = {10.1103/PhysRevLett.23.880}
}

@article{Beane:2018oxh,
    author = "Beane, Silas R. and Kaplan, David B. and Klco, Natalie and Savage, Martin J.",
    title = "{Entanglement Suppression and Emergent Symmetries of Strong Interactions}",
    eprint = "1812.03138",
    archivePrefix = "arXiv",
    primaryClass = "nucl-th",
    reportNumber = "INT-PUB-18-056, NT@UW-18-19",
    doi = "10.1103/PhysRevLett.122.102001",
    journal = "Phys. Rev. Lett.",
    volume = "122",
    pages = "102001",
    year = "2019"
}

@article{froissart1961asymptotic,
  title="{Asymptotic behavior and subtractions in the Mandelstam representation}",
  author={Froissart, Marcel},
  journal={Physical Review},
  volume={123},
  pages={1053},
  year={1961},
  publisher={APS}
}

@article{PhysRev.129.1432,
  title = {Unitarity and High-Energy Behavior of Scattering Amplitudes},
  author = {Martin, A.},
  journal = {Phys. Rev.},
  volume = {129},
  issue = {3},
  pages = {1432},
  year = {1963},
  publisher = {American Physical Society},
  doi = {10.1103/PhysRev.129.1432}
}

@article{Lukaszuk:1967zz,
    author = "Lukaszuk, L. and Martin, A.",
    title = "{Absolute upper bounds for pi pi scattering}",
    doi = "10.1007/BF02739279",
    journal = "Nuovo Cim. A",
    volume = "52",
    pages = "122",
    year = "1967"
}

@article{Adams:2006sv,
    author = "Adams, Allan and Arkani-Hamed, Nima and Dubovsky, Sergei and Nicolis, Alberto and Rattazzi, Riccardo",
    title = "{Causality, analyticity and an IR obstruction to UV completion}",
    eprint = "hep-th/0602178",
    archivePrefix = "arXiv",
    reportNumber = "CERN-PH-TH-2006-033, HUTP-06-A0005",
    doi = "10.1088/1126-6708/2006/10/014",
    journal = "JHEP",
    volume = "10",
    pages = "014",
    year = "2006"
}

@article{Jenkins:2006ia,
    author = "Jenkins, Alejandro and O'Connell, Donal",
    title = "{The Story of ${\cal O}$: Positivity constraints in effective field theories}",
    eprint = "hep-th/0609159",
    archivePrefix = "arXiv",
    reportNumber = "CALT-68-2607, MIT-CTP-3764",
    year = "2006"
}

@article{Dvali:2012zc,
    author = "Dvali, Gia and Franca, Andre and Gomez, Cesar",
    title = "{Road Signs for UV-Completion}",
    eprint = "1204.6388",
    archivePrefix = "arXiv",
    primaryClass = "hep-th",
    year = "2012"
}

@article{Remmen:2019cyz,
    author = "Remmen, Grant N. and Rodd, Nicholas L.",
    title = "{Consistency of the Standard Model Effective Field Theory}",
    eprint = "1908.09845",
    archivePrefix = "arXiv",
    primaryClass = "hep-ph",
    doi = "10.1007/JHEP12(2019)032",
    journal = "JHEP",
    volume = "12",
    pages = "032",
    year = "2019"
}

@article{Remmen:2020vts,
    author = "Remmen, Grant N. and Rodd, Nicholas L.",
    title = "{Flavor Constraints from Unitarity and Analyticity}",
    eprint = "2004.02885",
    archivePrefix = "arXiv",
    primaryClass = "hep-ph",
    doi = "10.1103/PhysRevLett.127.149901",
    journal = "Phys. Rev. Lett.",
    volume = "125",
    pages = "081601",
    year = "2020",
note           = "[Erratum: \href{https://doi.org/10.1103/PhysRevLett.127.149901}{{\it Phys. Rev. Lett.} {\bf 127} (2021) 149901}]"
}

@article{Remmen:2020uze,
    author = "Remmen, Grant N. and Rodd, Nicholas L.",
    title = "{Signs, spin, SMEFT: Sum rules at dimension six}",
    eprint = "2010.04723",
    archivePrefix = "arXiv",
    primaryClass = "hep-ph",
    doi = "10.1103/PhysRevD.105.036006",
    journal = "Phys. Rev. D",
    volume = "105",
    pages = "036006",
    year = "2022"
}

@article{Cheung:2020uts,
    author = "Cheung, Clifford and Remmen, Grant N.",
    title = "{Entanglement and the double copy}",
    eprint = "2002.10470",
    archivePrefix = "arXiv",
    primaryClass = "hep-th",
    reportNumber = "CALT-TH-2020-003",
    doi = "10.1007/JHEP05(2020)100",
    journal = "JHEP",
    volume = "05",
    pages = "100",
    year = "2020"
}

@article{Bellazzini:2015cra,
    author = "Bellazzini, Brando and Cheung, Clifford and Remmen, Grant N.",
    title = "{Quantum Gravity Constraints from Unitarity and Analyticity}",
    eprint = "1509.00851",
    archivePrefix = "arXiv",
    primaryClass = "hep-th",
    reportNumber = "CALT-TH-2015-044, SACLAY-T15-161",
    doi = "10.1103/PhysRevD.93.064076",
    journal = "Phys. Rev. D",
    volume = "93",
    pages = "064076",
    year = "2016"
}

@article{Cheung:2023hkq,
    author = "Cheung, Clifford and He, Temple and Sivaramakrishnan, Allic",
    title = "{Entropy growth in perturbative scattering}",
    eprint = "2304.13052",
    archivePrefix = "arXiv",
    primaryClass = "hep-th",
    reportNumber = "CALT-TH 2023-009",
    doi = "10.1103/PhysRevD.108.045013",
    journal = "Phys. Rev. D",
    volume = "108",
    pages = "045013",
    year = "2023"
}

@article{Fan:2017hcd,
    author = "Fan, Jinbo and Deng, Yanbin and Huang, Yong-Chang",
    title = "{Variation of entanglement entropy and mutual information in fermion-fermion scattering}",
    eprint = "1703.07911",
    archivePrefix = "arXiv",
    primaryClass = "hep-th",
    doi = "10.1103/PhysRevD.95.065017",
    journal = "Phys. Rev. D",
    volume = "95",
    pages = "065017",
    year = "2017"
}

@article{Peschanski:2019yah,
    author = "Peschanski, Robi and Seki, Shigenori",
    title = "{Evaluation of Entanglement Entropy in High Energy Elastic Scattering}",
    eprint = "1906.09696",
    archivePrefix = "arXiv",
    primaryClass = "hep-th",
    doi = "10.1103/PhysRevD.100.076012",
    journal = "Phys. Rev. D",
    volume = "100",
    pages = "076012",
    year = "2019"
}

@article{Peschanski,
    author = "Peschanski, Robi and Seki, Shigenori",
    title = "{Entanglement Entropy of Scattering Particles}",
    eprint = "1602.00720",
    archivePrefix = "arXiv",
    primaryClass = "hep-th",
    doi = "10.1016/j.physletb.2016.04.063",
    journal = "Phys. Lett. B",
    volume = "758",
    pages = "89",
    year = "2016"
}

@article{Balasubramanian:2011wt,
    author = "Balasubramanian, Vijay and McDermott, Michael B. and Van Raamsdonk, Mark",
    title = "{Momentum-space entanglement and renormalization in quantum field theory}",
    eprint = "1108.3568",
    archivePrefix = "arXiv",
    primaryClass = "hep-th",
    reportNumber = "UPR-1233-T",
    doi = "10.1103/PhysRevD.86.045014",
    journal = "Phys. Rev. D",
    volume = "86",
    pages = "045014",
    year = "2012"
}

@article{Costa:2022bvs,
    author = "Costa, Matheus H. Martins and Brink, Jeroen van den and Nogueira, Flavio S. and Krein, Gast\~ao I.",
    title = "{Momentum space entanglement from the Wilsonian effective action}",
    eprint = "2207.12103",
    archivePrefix = "arXiv",
    primaryClass = "hep-th",
    doi = "10.1103/PhysRevD.106.065024",
    journal = "Phys. Rev. D",
    volume = "106",
    pages = "065024",
    year = "2022"
}

@article{Rosenhaus_2014,
    author = "Rosenhaus, Vladimir and Smolkin, Michael",
    title = "{Entanglement Entropy: A Perturbative Calculation}",
    eprint = "1403.3733",
    archivePrefix = "arXiv",
    primaryClass = "hep-th",
    doi = "10.1007/JHEP12(2014)179",
    journal = "JHEP",
    volume = "12",
    pages = "179",
    year = "2014"
}

@article{Seki:2014cgq,
    author = "Seki, Shigenori and Park, I. Y. and Sin, Sang-Jin",
    title = "{Variation of Entanglement Entropy in Scattering Process}",
    eprint = "1412.7894",
    archivePrefix = "arXiv",
    primaryClass = "hep-th",
    doi = "10.1016/j.physletb.2015.02.028",
    journal = "Phys. Lett. B",
    volume = "743",
    pages = "147",
    year = "2015"
}

@article{Kosower:2018adc,
    author = "Kosower, David A. and Maybee, Ben and O'Connell, Donal",
    title = "{Amplitudes, Observables, and Classical Scattering}",
    eprint = "1811.10950",
    archivePrefix = "arXiv",
    primaryClass = "hep-th",
    doi = "10.1007/JHEP02(2019)137",
    journal = "JHEP",
    volume = "02",
    pages = "137",
    year = "2019"
}

@article{Low:2021ufv,
    author = "Low, Ian and Mehen, Thomas",
    title = "{Symmetry from entanglement suppression}",
    eprint = "2104.10835",
    archivePrefix = "arXiv",
    primaryClass = "hep-th",
    doi = "10.1103/PhysRevD.104.074014",
    journal = "Phys. Rev. D",
    volume = "104",
    pages = "074014",
    year = "2021"
}

@article{Arkani-Hamed:2016rak,
    author = "Arkani-Hamed, Nima and Rodina, Laurentiu and Trnka, Jaroslav",
    title = "{Locality and Unitarity of Scattering Amplitudes from Singularities and Gauge Invariance}",
    eprint = "1612.02797",
    archivePrefix = "arXiv",
    primaryClass = "hep-th",
    doi = "10.1103/PhysRevLett.120.231602",
    journal = "Phys. Rev. Lett.",
    volume = "120",
    pages = "231602",
    year = "2018"
}

@article{Arkani-Hamed:2023lbd,
    author = "Arkani-Hamed, N. and Frost, H. and Salvatori, G. and Plamondon, P-G. and Thomas, H.",
    title = "{All Loop Scattering as a Counting Problem}",
    eprint = "2309.15913",
    archivePrefix = "arXiv",
    primaryClass = "hep-th",
    year = "2023"
}

@article{Cheung:2016wjt,
    author = "Cheung, Clifford and Remmen, Grant N.",
    title = "{Positivity of Curvature-Squared Corrections in Gravity}",
    eprint = "1608.02942",
    archivePrefix = "arXiv",
    primaryClass = "hep-th",
    reportNumber = "CALT-TH-2016-018",
    doi = "10.1103/PhysRevLett.118.051601",
    journal = "Phys. Rev. Lett.",
    volume = "118",
    pages = "051601",
    year = "2017"
}

@article{Remmen:2022orj,
    author = "Remmen, Grant N. and Rodd, Nicholas L.",
    title = "{Spinning sum rules for the dimension-six SMEFT}",
    eprint = "2206.13524",
    archivePrefix = "arXiv",
    primaryClass = "hep-ph",
    reportNumber = "CERN-TH-2022-105",
    doi = "10.1007/JHEP09(2022)030",
    journal = "JHEP",
    volume = "09",
    pages = "030",
    year = "2022"
}

@article{Trott:2020ebl,
    author = "Trott, Timothy",
    title = "{Causality, unitarity and symmetry in effective field theory}",
    eprint = "2011.10058",
    archivePrefix = "arXiv",
    primaryClass = "hep-ph",
    doi = "10.1007/JHEP07(2021)143",
    journal = "JHEP",
    volume = "07",
    pages = "143",
    year = "2021"
}

@article{Arkani-Hamed:2021ajd,
    author = "Arkani-Hamed, Nima and Huang, Yu-tin and Liu, Jin-Yu and Remmen, Grant N.",
    title = "{Causality, unitarity, and the weak gravity conjecture}",
    eprint = "2109.13937",
    archivePrefix = "arXiv",
    primaryClass = "hep-th",
    doi = "10.1007/JHEP03(2022)083",
    journal = "JHEP",
    volume = "03",
    pages = "083",
    year = "2022"
}

@article{Zhang:2020jyn,
    author = "Zhang, Cen and Zhou, Shuang-Yong",
    title = "{Convex Geometry Perspective on the (Standard Model) Effective Field Theory Space}",
    eprint = "2005.03047",
    archivePrefix = "arXiv",
    primaryClass = "hep-ph",
    reportNumber = "USTC-ICTS/PCFT-20-14",
    doi = "10.1103/PhysRevLett.125.201601",
    journal = "Phys. Rev. Lett.",
    volume = "125",
    pages = "201601",
    year = "2020"
}

@article{Hayden:2011ag,
    author = "Hayden, Patrick and Headrick, Matthew and Maloney, Alexander",
    title = "{Holographic Mutual Information is Monogamous}",
    eprint = "1107.2940",
    archivePrefix = "arXiv",
    primaryClass = "hep-th",
    reportNumber = "BRX-TH-638",
    doi = "10.1103/PhysRevD.87.046003",
    journal = "Phys. Rev. D",
    volume = "87",
    pages = "046003",
    year = "2013"
}

@article{Aoude:2020mlg,
    author = "Aoude, Rafael and Chung, Ming-Zhi and Huang, Yu-tin and Machado, Camila S. and Tam, Man-Kuan",
    title = "{Silence of Binary Kerr Black Holes}",
    eprint = "2007.09486",
    archivePrefix = "arXiv",
    primaryClass = "hep-th",
    doi = "10.1103/PhysRevLett.125.181602",
    journal = "Phys. Rev. Lett.",
    volume = "125",
    pages = "181602",
    year = "2020"
}

@article{Chen:2021huj,
    author = "Chen, Bo-Ting and Chung, Ming-Zhi and Huang, Yu-tin and Tam, Man Kuan",
    title = "{Minimal spin deflection of Kerr-Newman and supersymmetric black hole}",
    eprint = "2106.12518",
    archivePrefix = "arXiv",
    primaryClass = "hep-th",
    doi = "10.1007/JHEP10(2021)011",
    journal = "JHEP",
    volume = "10",
    pages = "011",
    year = "2021"
}

@book{nielsen00,
  author = {Nielsen, Michael A. and Chuang, Isaac L.},
  publisher = {Cambridge University Press},
  title = {Quantum Computation and Quantum Information},
  year = 2000
}

@article{ATLAS:2023fsd,
    author = "Aad, Georges and others",
    collaboration = "ATLAS",
    title = "{Observation of quantum entanglement in top-quark pairs using the ATLAS detector}",
    eprint = "2311.07288",
    archivePrefix = "arXiv",
    primaryClass = "hep-ex",
    reportNumber = "CERN-EP-2023-230",
    year = "2023"
}

@article{Severi:2021cnj,
    author = "Severi, Claudio and Boschi, Cristian Degli Esposti and Maltoni, Fabio and Sioli, Maximiliano",
    title = "{Quantum tops at the LHC: from entanglement to Bell inequalities}",
    eprint = "2110.10112",
    archivePrefix = "arXiv",
    primaryClass = "hep-ph",
    doi = "10.1140/epjc/s10052-022-10245-9",
    journal = "Eur. Phys. J. C",
    volume = "82",
    pages = "285",
    year = "2022"
}

@article{Aguilar-Saavedra:2022uye,
    author = "Aguilar-Saavedra, J. A. and Casas, J. A.",
    title = "{Improved tests of entanglement and Bell inequalities with LHC tops}",
    eprint = "2205.00542",
    archivePrefix = "arXiv",
    primaryClass = "hep-ph",
    reportNumber = "IFT-UAM/CSIC-22-45",
    doi = "10.1140/epjc/s10052-022-10630-4",
    journal = "Eur. Phys. J. C",
    volume = "82",
    pages = "666",
    year = "2022"
}

@article{Afik:2020onf,
    author = "Afik, Yoav and de Nova, Juan Ram\'on Mu\~noz",
    title = "{Entanglement and quantum tomography with top quarks at the LHC}",
    eprint = "2003.02280",
    archivePrefix = "arXiv",
    primaryClass = "quant-ph",
    doi = "10.1140/epjp/s13360-021-01902-1",
    journal = "Eur. Phys. J. Plus",
    volume = "136",
    pages = "907",
    year = "2021"
}

@article{Aoude:2022imd,
    author = "Aoude, Rafael and Madge, Eric and Maltoni, Fabio and Mantani, Luca",
    title = "{Quantum SMEFT tomography: Top quark pair production at the LHC}",
    eprint = "2203.05619",
    archivePrefix = "arXiv",
    primaryClass = "hep-ph",
    reportNumber = "CP3-22-14",
    doi = "10.1103/PhysRevD.106.055007",
    journal = "Phys. Rev. D",
    volume = "106",
    pages = "055007",
    year = "2022"
}

@article{Fabbrichesi:2022ovb,
    author = "Fabbrichesi, Marco and Floreanini, Roberto and Gabrielli, Emidio",
    title = "{Constraining new physics in entangled two-qubit systems: top-quark, tau-lepton and photon pairs}",
    eprint = "2208.11723",
    archivePrefix = "arXiv",
    primaryClass = "hep-ph",
    doi = "10.1140/epjc/s10052-023-11307-2",
    journal = "Eur. Phys. J. C",
    volume = "83",
    pages = "162",
    year = "2023"
}

@article{Severi:2022qjy,
    author = "Severi, Claudio and Vryonidou, Eleni",
    title = "{Quantum entanglement and top spin correlations in SMEFT at higher orders}",
    eprint = "2210.09330",
    archivePrefix = "arXiv",
    primaryClass = "hep-ph",
    doi = "10.1007/JHEP01(2023)148",
    journal = "JHEP",
    volume = "01",
    pages = "148",
    year = "2023"
}

@article{Barr:2021zcp,
    author = "Barr, Alan J.",
    title = "{Testing Bell inequalities in Higgs boson decays}",
    eprint = "2106.01377",
    archivePrefix = "arXiv",
    primaryClass = "hep-ph",
    doi = "10.1016/j.physletb.2021.136866",
    journal = "Phys. Lett. B",
    volume = "825",
    pages = "136866",
    year = "2022"
}

@article{Ashby-Pickering:2022umy,
    author = "Ashby-Pickering, Rachel and Barr, Alan J. and Wierzchucka, Agnieszka",
    title = "{Quantum state tomography, entanglement detection and Bell violation prospects in weak decays of massive particles}",
    eprint = "2209.13990",
    archivePrefix = "arXiv",
    primaryClass = "quant-ph",
    doi = "10.1007/JHEP05(2023)020",
    journal = "JHEP",
    volume = "05",
    pages = "020",
    year = "2023"
}

@article{Dutta:2019gen,
    author = "Dutta, Souvik and Faulkner, Thomas",
    title = "{A canonical purification for the entanglement wedge cross-section}",
    eprint = "1905.00577",
    archivePrefix = "arXiv",
    primaryClass = "hep-th",
    doi = "10.1007/JHEP03(2021)178",
    journal = "JHEP",
    volume = "03",
    pages = "178",
    year = "2021"
}

@article{Bao:2021vyq,
    author = "Bao, Ning and Chatwin-Davies, Aidan and Remmen, Grant N.",
    title = "{Entanglement wedge cross section inequalities from replicated geometries}",
    eprint = "2106.02640",
    archivePrefix = "arXiv",
    primaryClass = "hep-th",
    doi = "10.1007/JHEP07(2021)113",
    journal = "JHEP",
    volume = "07",
    pages = "113",
    year = "2021"
}

@article{Lindblad,
author = "Lindblad, G.",
title = "Expectations and entropy inequalities for finite quantum systems",
journal = "Commun.Math. Phys.",
volume = "39", 
pages = "111",
year = "1974",
doi = "10.1007/BF01608390"
}

@article{Evenbly:2007hxg,
    author = "Evenbly, G. and Vidal, G.",
    title = "{Algorithms for entanglement renormalization}",
    eprint = "0707.1454",
    archivePrefix = "arXiv",
    primaryClass = "cond-mat.str-el",
    doi = "10.1103/PhysRevB.79.144108",
    journal = "Phys. Rev. B",
    volume = "79",
    pages = "144108",
    year = "2009"
}

@article{Bao:2015uaa,
    author = "Bao, Ning and Cao, ChunJun and Carroll, Sean M. and Chatwin-Davies, Aidan and Hunter-Jones, Nicholas and Pollack, Jason and Remmen, Grant N.",
    title = "{Consistency conditions for an AdS multiscale entanglement renormalization ansatz correspondence}",
    eprint = "1504.06632",
    archivePrefix = "arXiv",
    primaryClass = "hep-th",
    reportNumber = "CALT-TH-2015-015",
    doi = "10.1103/PhysRevD.91.125036",
    journal = "Phys. Rev. D",
    volume = "91",
    pages = "125036",
    year = "2015"
}

@article{Freytsis:2022aho,
    author = "Freytsis, Marat and Kumar, Soubhik and Remmen, Grant N. and Rodd, Nicholas L.",
    title = "{Multifield positivity bounds for inflation}",
    eprint = "2210.10791",
    archivePrefix = "arXiv",
    primaryClass = "hep-th",
    reportNumber = "CERN-TH-2022-160",
    doi = "10.1007/JHEP09(2023)041",
    journal = "JHEP",
    volume = "09",
    pages = "041",
    year = "2023"
}

@article{Carena:2023vjc,
    author = "Carena, Marcela and Low, Ian and Wagner, Carlos E. M. and Xiao, Ming-Lei",
    title = "{Entanglement Suppression, Enhanced Symmetry and a Standard-Model-like Higgs Boson}",
    eprint = "2307.08112",
    archivePrefix = "arXiv",
    primaryClass = "hep-ph",
    reportNumber = "FERMILAB-PUB-23-269-T",
    year = "2023"
}

@article{Bellazzini:2016xrt,
    author = "Bellazzini, Brando",
    title = "{Softness and amplitudes' positivity for spinning particles}",
    eprint = "1605.06111",
    archivePrefix = "arXiv",
    primaryClass = "hep-th",
    reportNumber = "SACLAY-T16-038",
    doi = "10.1007/JHEP02(2017)034",
    journal = "JHEP",
    volume = "02",
    pages = "034",
    year = "2017"
}

@article{Arkani-Hamed:2020blm,
    author = "Arkani-Hamed, Nima and Huang, Tzu-Chen and Huang, Yu-tin",
    title = "{The EFT-Hedron}",
    eprint = "2012.15849",
    archivePrefix = "arXiv",
    primaryClass = "hep-th",
    reportNumber = "NCTS-TH/2014, CALT-TH 2020-061",
    doi = "10.1007/JHEP05(2021)259",
    journal = "JHEP",
    volume = "05",
    pages = "259",
    year = "2021"
}

@article{Cheung:2014ega,
    author = "Cheung, Clifford and Remmen, Grant N.",
    title = "{Infrared Consistency and the Weak Gravity Conjecture}",
    eprint = "1407.7865",
    archivePrefix = "arXiv",
    primaryClass = "hep-th",
    reportNumber = "CALT-TH-2014-146",
    doi = "10.1007/JHEP12(2014)087",
    journal = "JHEP",
    volume = "12",
    pages = "087",
    year = "2014"
}

@article{Hamada:2018dde,
    author = "Hamada, Yuta and Noumi, Toshifumi and Shiu, Gary",
    title = "{Weak Gravity Conjecture from Unitarity and Causality}",
    eprint = "1810.03637",
    archivePrefix = "arXiv",
    primaryClass = "hep-th",
    reportNumber = "CCTP-2018-12, ITCP-IPP 2018/9, KOBE-COSMO-18-08, MAD-TH-18-05",
    doi = "10.1103/PhysRevLett.123.051601",
    journal = "Phys. Rev. Lett.",
    volume = "123",
    pages = "051601",
    year = "2019"
}

@article{Nicolis:2009qm,
    author = "Nicolis, Alberto and Rattazzi, Riccardo and Trincherini, Enrico",
    title = "{Energy's and amplitudes' positivity}",
    eprint = "0912.4258",
    archivePrefix = "arXiv",
    primaryClass = "hep-th",
    doi = "10.1007/JHEP05(2010)095",
    journal = "JHEP",
    volume = "05",
    pages = "095",
    year = "2010",
note           = "[Erratum: \href{https://doi.org/10.1007/JHEP11(2011)128}{{\it JHEP} {\bf 11} (2011) 128}]"
}

@article{Sinha:2020win,
    author = "Sinha, Aninda and Zahed, Ahmadullah",
    title = "{Crossing Symmetric Dispersion Relations in Quantum Field Theories}",
    eprint = "2012.04877",
    archivePrefix = "arXiv",
    primaryClass = "hep-th",
    doi = "10.1103/PhysRevLett.126.181601",
    journal = "Phys. Rev. Lett.",
    volume = "126",
    pages = "181601",
    year = "2021"
}

@article{Sinha:2022crx,
    author = "Sinha, Aninda and Zahed, Ahmadullah",
    title = "{Bell inequalities in 2-2 scattering}",
    eprint = "2212.10213",
    archivePrefix = "arXiv",
    primaryClass = "hep-th",
    doi = "10.1103/PhysRevD.108.025015",
    journal = "Phys. Rev. D",
    volume = "108",
    pages = "025015",
    year = "2023"
}

@article{Cheung:2025nhw,
    author = "Cheung, Clifford and Remmen, Grant N.",
    title = "{Multipositivity bounds for scattering amplitudes}",
    eprint = "2505.05553",
    archivePrefix = "arXiv",
    primaryClass = "hep-th",
    reportNumber = "CALT-TH 2025-010",
    doi = "10.1103/wt4x-2149",
    journal = "Phys. Rev. D",
    volume = "112",
    pages = "016017",
    year = "2025"
}

@article{Remmen:2024hry,
    author = "Remmen, Grant N. and Rodd, Nicholas L.",
    title = "{Positively Identifying HEFT or SMEFT}",
    eprint = "2412.07827",
    archivePrefix = "arXiv",
    primaryClass = "hep-ph",
    month = "12",
    year = "2024"
}

\end{document}